\documentclass[journal]{IEEEtran}

\usepackage[utf8]{inputenc}
\usepackage[T1]{fontenc}
\usepackage{amssymb}
\usepackage{lipsum}
\usepackage{amsthm}
\usepackage{subfigure}
\usepackage{gensymb}
\usepackage{textcomp}
\usepackage{multirow}
\usepackage{makecell}
\usepackage{soul,framed} 
\usepackage{url}
\usepackage[table]{xcolor}
\usepackage{amsmath,amsfonts}
\usepackage{graphicx}
\usepackage{dcolumn}
\usepackage{pict2e}
\usepackage{epstopdf}
\usepackage{arydshln}
\usepackage[mathscr]{euscript}
\usepackage{algorithmic}
\usepackage{algorithm}
\usepackage{array}
\usepackage{textcomp}
\usepackage{verbatim}
\usepackage{graphicx}
\usepackage{stfloats}
\usepackage{float}
\usepackage{pifont}
\usepackage{colortbl}
\usepackage{pgf} 

\definecolor{bluerevision}{RGB}{0,112,192}

\makeatletter
 \let\old@ps@headings\ps@headings
 \let\old@ps@IEEEtitlepagestyle\ps@IEEEtitlepagestyle
 \def\confheader#1{%
 \def\ps@headings{%
 \old@ps@headings%
 \def\@oddhead{\strut\hfill#1\hfill\strut}%
 \def\@evenhead{\strut\hfill#1\hfill\strut}%
 }%
 \def\ps@IEEEtitlepagestyle{%
 \old@ps@IEEEtitlepagestyle%
 \def\@oddhead{\strut\hfill#1\hfill\strut}%
 \def\@evenhead{\strut\hfill#1\hfill\strut}%
 }%
 \ps@headings%
 }
 \makeatother
\confheader{%
This work has been accepted for publication in \textit{Results in Engineering.} DOI: 10.1016/j.rineng.2026.113002}

\begin{document}

\title{On the LEO Satellite Constellation Design for North Atlantic Coverage}

\author{Alejandro Ramírez-Arroyo, Miguel Villanueva-Fernández, Preben Mogensen

\thanks{\textit{Corresponding author: Alejandro Ramírez-Arroyo.}}

\thanks{Alejandro Ramírez-Arroyo, Miguel Villanueva-Fernández and Preben Mogensen are with the Department of Electronic Systems, Aalborg University (AAU), 9220 Aalborg, Denmark (e-mail: araar@es.aau.dk, mvf@es.aau.dk, pm@es.aau.dk).}}

\markboth{Ramírez-Arroyo \MakeLowercase{\textit{et al.}}: On the LEO Satellite Constellation Design for North Atlantic Coverage}%
{Ramírez-Arroyo \MakeLowercase{\textit{et al.}}: On the LEO Satellite Constellation Design for North Atlantic Coverage}

\IEEEpubid{ }

\maketitle

\begin{abstract}
Low Earth Orbit (LEO) satellite constellations are emerging as a key component of non-terrestrial networks due to their low-latency and high-capacity communication capabilities. However, satellites in these orbits are characterized by a small coverage footprint and high orbital velocity compared to those in higher orbits. This results in constantly changing and dynamic constellations that require smart design of orbital parameters to ensure continuous coverage. Existing constellation deployments are typically optimized either for low- and mid-latitude regions or for full polar coverage, leaving high-latitude regional scenarios such as the North Atlantic insufficiently explored. This work provides insights into the key characteristics associated with the deployment of satellites in LEO for North Atlantic coverage. Therefore, we investigate through a parametric study the individual influence on how constellation inclination, minimum elevation angle, altitude, and satellite footprint affect visibility probability, revisit time, path loss, and coverage continuity. Results show that the minimum elevation angle is a critical design parameter since a Walker Delta constellation with 64 satellites at 1000 km altitude can provide continuous visibility above 55$\degree$N for elevations below 20$\degree$, whereas visibility probability degrades drastically for larger elevation angles. Similarly, inclinations above approximately 70$\degree$ are required to achieve robust North Atlantic coverage with medium-size constellations. Thus, these results provide practical guidelines on how a satellite constellation should be designed to achieve an efficient deployment with a focus on visibility coverage over the North Atlantic, targeting maritime, aviation, and Arctic connectivity scenarios.
\end{abstract}

\begin{IEEEkeywords}
NTN, Satellite, LEO, Design, Coverage, Inclination 
\end{IEEEkeywords}

\maketitle

\section{Introduction}

The development of wireless communication technologies has had ubiquitous high-speed connectivity as one of its main goals. Even though cellular terrestrial networks give coverage to a high percentage of the land in many countries, there are inherent limitations in specific regions due to the difficulty or the high costs of setting up infrastructure~\cite{6G_NTN, toward_6G_NTN}. Therefore, Geostationary Earth Orbit (GEO) satellites were among the first non-terrestrial network (NTN) solutions to provide global coverage and service in remote areas~\cite{GEO}. The main characteristic of this type of orbit is that they are located in the equatorial plane at an altitude of 35,876~km above sea level. This orbit allows for an orbital period equal to the Earth’s rotation period, thereby generating a stationary view from the perspective of a user on Earth. The advantage of these orbits is the stationary position of the satellites relative to Earth, as well as the fact that, due to the high altitude, most of the Earth can be covered with just three satellites. Nevertheless, their position in the equatorial plane prevents coverage at extreme latitudes, i.e., the Earth’s poles, and their high altitude limits the effective throughput and the minimum latency of communications, which is lower bounded by approximately 480 ms due to the round-trip time in two-way communications between Earth and satellites in GEO orbits. For these reasons, another approach proposed for establishing non-terrestrial communications is based on Low Earth Orbit (LEO) constellations~\cite{architecture_delta_star, LEO_Leyva_book}. These constellations are unique in that their orbits range from 160~km to 2,000~km, and more commonly between 500 km and 1,200~km in altitude. One consequence of this lower orbit is that they have an average speed of 7.8 km/s, which implies relative motion of the satellite with respect to the ground station on Earth, resulting in orbital periods of approximately 90 to 120~minutes. Such characteristics result in higher throughput due to reduced path loss, as well as lower latency due to the shorter range. However, the satellites' movement relative to Earth and their lower altitude, which reduces their footprint, require a greater number of satellites to provide continuous coverage of a given area, making the design of the constellation's orbital parameters a critical aspect depending on the use cases to be covered~\cite{LEO_Leyva}.

The growing demand for connectivity in recent years has led to the rise of LEO constellations on the market~\cite{evolution_NTN}. These networks cover a wide range of use cases, such as broadband connectivity in remote areas, IoT, and maritime or aviation traffic. Most of these constellations follow a deployment pattern based on the Walker Delta and Walker Star geometries~\cite{architecture_delta_star}. The former maximizes coverage at low inclinations, making it particularly useful for covering regions with low or mid-latitudes, where the majority of the global population is clustered~\cite{satellite-comparison}. The latter tends to provide global coverage since its orbits pass over the Earth’s poles, ensuring visibility across the entire globe. The design philosophy behind both geometries means that, in practice, Walker Delta tends to be deployed at inclinations below 55$\degree$, while Walker Star tends to be deployed at inclinations close to 90$\degree$. Therefore, the range of intermediate inclinations has been little studied, despite the existence of established and emerging use cases that show some interest in satellite constellations providing coverage in this range of latitudes. These include, among others, the previously mentioned air and maritime traffic, offshore platforms, and coverage of remote regions such as northern Canada, Greenland, and Svalbard, as well as emerging use cases like maritime surveillance in the Atlantic and Arctic regions, driven by the opening of new shipping routes as a consequence of ice melt.

Extensive analytical and simulation work has been conducted in the literature to model and optimize aspects related to the performance and design of non-terrestrial networks. Among others, Al Homssi~\textit{et al.}~\cite{uplink_hybrid_satellite} analytically examine the performance of hybrid satellite-terrestrial networks, evaluating the size of the constellation versus the density of the terrestrial infrastructure. They also propose a model for optimizing uplink performance based solely on the constellation altitude and the beamwidth of the satellite antennas~\cite{uplink_performance}. Al-Hourani \textit{et al.}~\cite{S2G_PL, coverage_dense_networks} analyze and model the path loss and coverage for non-terrestrial networks in dense networks for urban areas. Liang \textit{et al.}~\cite{phasing} study how the phasing between adjacent orbital planes affects the distances between satellites and maximize the minimum inter-satellite distance to reduce the likelihood of collisions in mega-constellations. Fastenbauer \textit{et al.}~\cite{elevation_angle_analysis} investigate the impact of the elevation angle on the satellite footprint for multi-beam systems and how it affects beam design and the signal-to-interference-and-noise ratio. Jeon \textit{et al.}~\cite{Voronoi_APC} study coverage continuity in non-terrestrial networks based on the analysis of Voronoi diagrams and access profile, constellation pattern, and coverage timeline decompositions. At the system level, several efforts have also been made to characterize the operation of satellite networks. For instance, Leyva \textit{et al.}~\cite{LEO_Leyva} conduct a study of how to integrate the coverage of LEO constellations into the architecture of 5G systems, while Dakic \textit{et al.}~\cite{RAAN} investigate energy efficiency and end-to-end delay based on the choice of inter-satellite links.

This paper provides an overview of the key parameters that model Walker Delta and Walker Star orbits for LEO satellite constellations, and their impact on coverage conditions of a non-terrestrial satellite network. In particular, the work is motivated by the coverage limitations of existing constellation designs in mid- and high-latitude regions, where low-inclination constellations provide insufficient coverage while polar constellations may over-cover non-priority areas. This creates a critical intermediate region in which constellation geometry is not fully optimized for either scenario. To address this challenge, the paper focuses on the parametric study of visibility probability and revisit time as a function of parameters such as minimum elevation angle and constellation inclination, with special emphasis on deployments with ground stations in mid- and high-latitude regions where emerging use cases may be of particular relevance. Recent studies have also analyzed LEO constellation performance in Arctic and other high-latitude regions. For instance, Maleki \textit{et al.}~\cite{ICC_polar_LEO} develop a stochastic-geometry framework for Arctic and subarctic coverage, considering user latitudes between 60$^\circ$N and 90$^\circ$N. Their model assumes satellites deployed in circular polar orbits with stochastic positions along the orbital tracks and derives analytical expressions for coverage probability and ergodic capacity, which are subsequently validated through Monte Carlo simulations. In contrast, the present study focuses on the deterministic orbital geometry of Walker Delta-type constellations for regional coverage over the North Atlantic. As previously stated, the analysis explicitly considers the constellation inclination, number of satellites, number of orbital planes, phasing parameter, altitude, and minimum elevation angle, and evaluates their individual influence on visibility probability and revisit time through simulations. Furthermore, the geographical distribution of visibility is explicitly analysed over the North Atlantic, together with the associated coverage footprint. Thus, while both studies address LEO coverage in high-latitude regions, this work complements the stochastic analytical treatment in~\cite{ICC_polar_LEO} by focusing on the temporal and spatial behavior of Walker Delta-type constellation configurations, including visibility probability and revisit time. In addition, the paper addresses other important factors to consider when designing constellations, such as the effect of satellite altitude on the footprint, path loss, and Line-of-Sight conditions when establishing communications. Finally, the work provides an overview of the current status of deployed LEO constellations and those currently being deployed, along with their characteristics and use cases. By integrating the previous ideas, this works aims to improve the understanding on the fundamental parameters underlying satellite constellation design and to systematically assess the influence of individual orbital parameter variations on the resulting ground coverage characteristics. Therefore, rather than replacing existing optimization-based constellation design tools, which are well-covered given the previous state-of-the-art, the objective of this study is to complement them by providing engineering insight into the influence of the main orbital design parameters on regional coverage performance.

The work is organized as follows. Section~II presents an overview of the current status of the main constellations deployed in LEO orbits, as well as some of those currently under development. Section~III introduces the basic concepts of coverage for LEO constellations, including their geometry, footprint, Line-of-Sight, and path loss. Section~IV analyzes the most relevant geometric factors for establishing coverage around the North Atlantic through a parametric study where each factor is considered separately, taking into account indicators such as visibility and revisit time across multiple geographical latitudes. Finally, Section~V summarizes the main conclusions of this study.

\begin{table*}[t]
\centering
\caption{Comparison of orbital parameters, design, and use cases for multiple LEO satellite communication constellations.\\Data as of January 2026.}
\label{tab:comparacion_tabla}
\resizebox{\textwidth}{!}{%
\begin{tabular}{||c||c|c|c|c|c|c|c||}
\hline
\hline
\textbf{Constellation} &
  \textbf{Operator} &
  \textbf{Status} &
  \textbf{N\textsubscript{sat}} &
  \textbf{Orbital Configuration} &
  \textbf{Altitude} &
  \textbf{Frequency Band} &
  \textbf{Use Cases} \\ \hline\hline
\textbf{Starlink} &
  SpaceX &
  Operational &
  $>$ 9000 &
  \begin{tabular}[c]{@{}c@{}}Multi-Shell\\ \textgreater 100 planes\\ 53$\degree$ inclination (most orbits)\\ Few orbits at high latitudes\\ ($\sim$70-90$\degree$ inclination)\end{tabular} &
  \begin{tabular}[c]{@{}c@{}}525-570 km\\ (typical)\end{tabular} &
  \begin{tabular}[c]{@{}c@{}}Ku-Band\\ Ka-Band\\ V-Band\\ E-Band\end{tabular} &
  \begin{tabular}[c]{@{}c@{}}- Global broadband connectivity\\ - Network resilience and backup\\ - Mobility support (aviation,\\ maritime and land)\end{tabular} \\ \hline
\textbf{OneWeb} &
  Eutelsat &
  Operational &
  648 &
  \begin{tabular}[c]{@{}c@{}}Polar Orbit (Walker Star)\\ 12 planes\\ 87.9$\degree$ inclination\end{tabular} &
  1200 km &
  \begin{tabular}[c]{@{}c@{}}Ku-Band\\ Ka-Band\end{tabular} &
  \begin{tabular}[c]{@{}c@{}}- Enterprise and government\\ broadband connectivity\\ - Network resilience and backup\\ - Mobility support (aviation,\\ maritime and land)\end{tabular} \\ \hline
\textbf{Iridium NEXT} &
  \begin{tabular}[c]{@{}c@{}}Iridium\\ Communications\end{tabular} &
  Operational &
  66 &
  \begin{tabular}[c]{@{}c@{}}Polar orbit (Walker Star)\\ 6 planes\\ 86.4$\degree$ inclination\end{tabular} &
  778 km &
  \begin{tabular}[c]{@{}c@{}}L-band\\ Ka-band\end{tabular} &
  \begin{tabular}[c]{@{}c@{}}- Global narrowband\\ communications (voice and data)\\ - Emergency services\\ (maritime, aviation, IoT)\end{tabular} \\ \hline
\textbf{GlobalStar} &
  GlobalStar &
  Operational &
  24 &
  \begin{tabular}[c]{@{}c@{}}Walker Delta\\ 8 planes\\ 52$\degree$ Inclination\end{tabular} &
  1414 km &
  \begin{tabular}[c]{@{}c@{}}C-Band\\ L-Band\\ S-Band\end{tabular} &
  \begin{tabular}[c]{@{}c@{}}- Global narrowband\\ communications (voice and data)\\ - Emergency services\\ (maritime, aviation, IoT)\end{tabular} \\ \hline
\textbf{Amazon LEO} &
  Amazon &
  \begin{tabular}[c]{@{}c@{}}Under\\ development\end{tabular} &
  \begin{tabular}[c]{@{}c@{}}3236\\ (Expected)\end{tabular} &
  \begin{tabular}[c]{@{}c@{}}Multi-Shell (3 orbital shells)\\ 98 planes\\ 33$\degree$, 42$\degree$ and 51.9$\degree$\end{tabular} &
  590, 610 and 630 km &
  \begin{tabular}[c]{@{}c@{}}Ku-Band\\ Ka-Band\end{tabular} &
  \begin{tabular}[c]{@{}c@{}}- Global broadband connectivity\\ - Integration with cloud services\\ and enterprise solutions\end{tabular} \\ \hline
\textbf{IRIS\textsuperscript{2}} &
  European Union &
  \begin{tabular}[c]{@{}c@{}}Under\\ development\end{tabular} &
  \begin{tabular}[c]{@{}c@{}}290\\ (Expected)\end{tabular} &
  \begin{tabular}[c]{@{}c@{}}Multi-shell\\ LEO (274 Sats.)\\ MEO (18 Sats.)\end{tabular} &
  \begin{tabular}[c]{@{}c@{}}1200 km (LEO)\\ 8000 km (MEO)\end{tabular} &
  \begin{tabular}[c]{@{}c@{}}Ku-Band\\ Ka-Band\\ K-Band\end{tabular} &
  \begin{tabular}[c]{@{}c@{}}- Secure governmental communications \\ - Resilient connectivity\\ for critical infrastructure\\ - Commercial broadband\\ services in Europe\end{tabular} \\ \hline\hline
\end{tabular}%
}
\end{table*}


\section{Current Status of LEO Constellations}

LEO constellations are currently the most widespread solution for non-terrestrial connectivity due to their high data-rate, low latency and increasingly low cost in the recent years. Several commercial LEO satellite communication constellations are compared in Table \ref{tab:comparacion_tabla}, with four different options currently in operation represented by Starlink, OneWeb, Iridium NEXT and GlobalStar; and two more under development with Amazon LEO and IRIS\textsuperscript{2}. However, not all serve the same purposes and use cases, as they have distinct characteristics. Therefore, this Section describes the specific features and purposes of each of them.

SpaceX owns the Starlink communication constellation, currently the largest in orbit, with over 9,000 operational satellites and a licensed capacity of up to 15,000 at the time of writing~\cite{starlink_latest}. To accommodate all satellites, more than a hundred planes in a multi-shell configuration are employed, at an inclination of 53$\degree$ for most orbits, although a limited number of shells at higher inclination have been deployed to improve coverage towards polar regions. Most of the satellites operate at a service altitude of 525 to 570 km, even though further planned deployments would bring down the minimum height to 355 km. They also make use of many different communication bands with Ku, Ka, V and E bands, which translates to a frequency range from 10.7 GHz to 86.0 GHz. All these characteristics make Starlink the current constellation with the highest possible throughput capacity and lowest latency due to the high number of satellites and their low orbit altitude, which enables the possibility of global broadband connectivity among other use cases, such as network resilience, backup and mobility support \cite{satellite-comparison}. However, with such use case as a primary goal, the coverage in less dense higher latitude regions is not a priority.

Oneweb is Eutelsat's offering, and represents the only alternative in operation to Starlink in terms of broadband connectivity. It has 648 satellites in a Walker Star polar orbit with 12 planes and a 87.9$\degree$ inclination at an approximate 1200 km altitude \cite{oneweb_fcc}. The lower total satellite count is compensated by a higher orbital altitude, which increases satellite footprint size and reduces handover frequency, a particularly relevant feature for oceanic and high-latitude services. However, this factor in combination with only using the Ku and Ka bands for communication leads to a lower overall throughput capacity for the services. Such limitations aim the same broadband connectivity services and use cases towards enterprises and governments instead of the general public. On the other hand, it is also a constellation better suited for total global coverage even in high latitude scenarios.

Iridium Communications is the operator of the Iridium NEXT, a constellation that represents a different design philosophy compared to the previous broadband-oriented LEO constellations, as the primary objective is global coverage reliability rather than high throughput. The constellation consists of 66 operational satellites deployed in a near-polar Walker Star configuration with six orbital planes at an inclination of 86.4$\degree$ at an approximate altitude of 780 km \cite{iridium_fcc}. Such an orbital configuration guarantees continuous global coverage, including polar and oceanic regions, with a relatively small number of satellites. The main operating frequencies lay in the L-band, with some supplementary Ka-band feeder links additionally used. These lower frequencies ensure lower susceptibility to atmospheric attenuation and severe environmental conditions, but at the cost of significantly lower data rates. Nevertheless, the constellation is widely used for safety-critical applications such as aviation, maritime communications, emergency services and Internet-of-Things (IoT) connectivity. Its near-polar inclination and extensive use of inter-satellite links make Iridium particularly well suited for uninterrupted service in high-latitude regions, including the North Atlantic.

Globalstar is another legacy LEO communication constellation currently in operation, although with more limited capabilities compared to Iridium NEXT.
It consists of 24 satellites deployed in a Walker Delta configuration with eight orbital planes at an inclination of 52$\degree$ and an orbital altitude of approximately 1,414 km \cite{globalstar_fcc}. Unlike polar constellations, this inclination provides reduced coverage at high latitudes, which limits service availability in regions closest to the poles. Globalstar primarily operates in the L and S bands, focusing on low data rate and voice services, emergency communications and IoT applications. The constellation does not rely on inter-satellite links, requiring continuous visibility of both user terminals and ground gateways, which further constrains coverage over oceanic regions. As a result, Globalstar is less suitable for continuous coverage when compared to polar-orbiting LEO systems.

Amazon’s LEO constellation, commonly referred to as Project Kuiper, is currently under development and is expected to deploy more than 3,200 satellites once fully operational \cite{kuiper_fcc}. The system is designed as a multi-shell constellation with orbits distributed across different altitudes, typically between 590 km and 630 km, and inclinations of 33$\degree$, 42$\degree$ and 51.9$\degree$. Operating primarily in the Ku and Ka bands, Amazon LEO aims to provide low-latency and high-capacity global broadband connectivity tightly integrated with cloud services and enterprise solutions. However, the absence of orbital shells close to the pole implies that service continuity at high latitudes, such as the North Atlantic region, is not a primary design objective.

\begin{figure*}[!b]
	\centering
	\subfigure{\includegraphics[width=0.8\columnwidth]{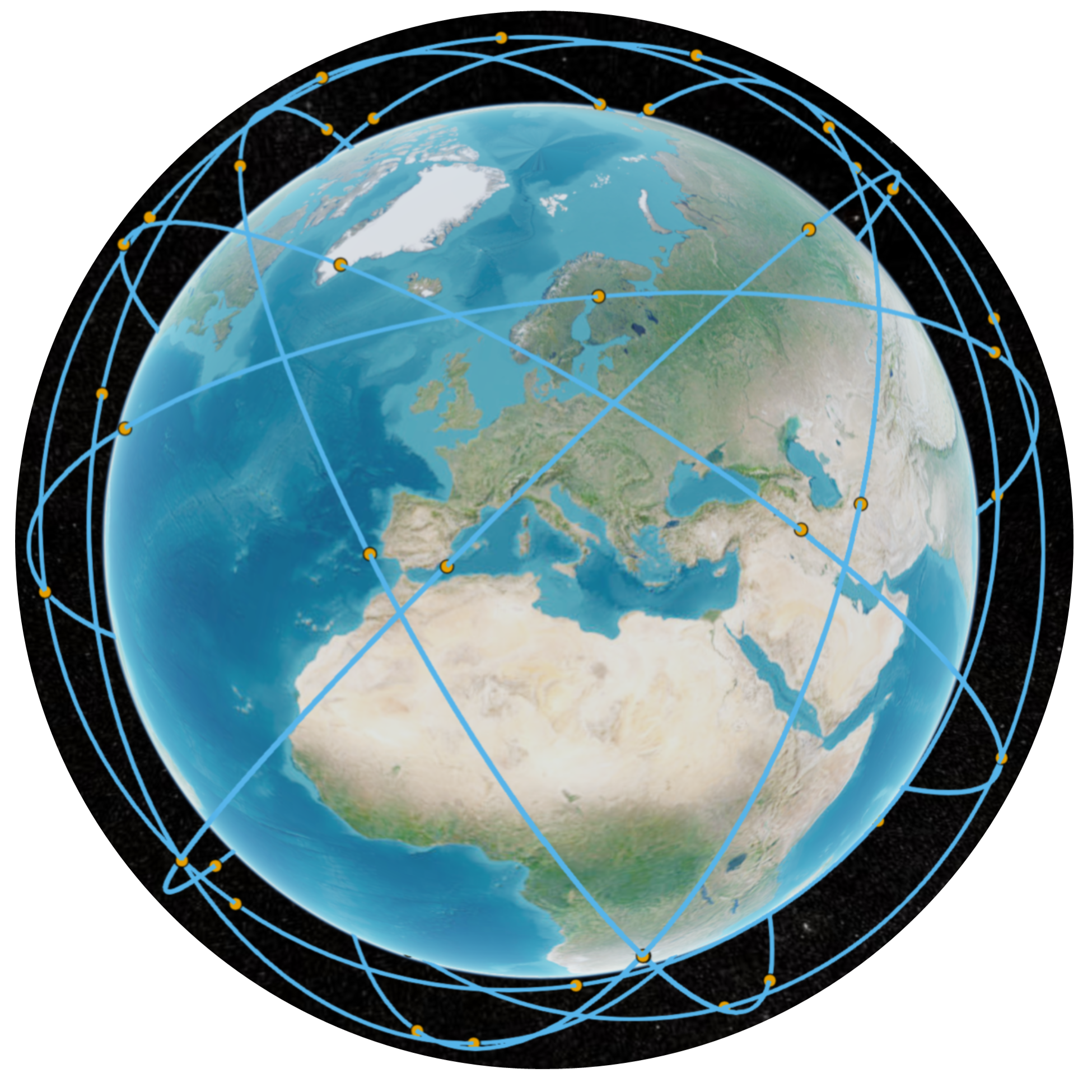}
	} \hspace{1.3cm}
    \subfigure{\includegraphics[width=0.8\columnwidth]{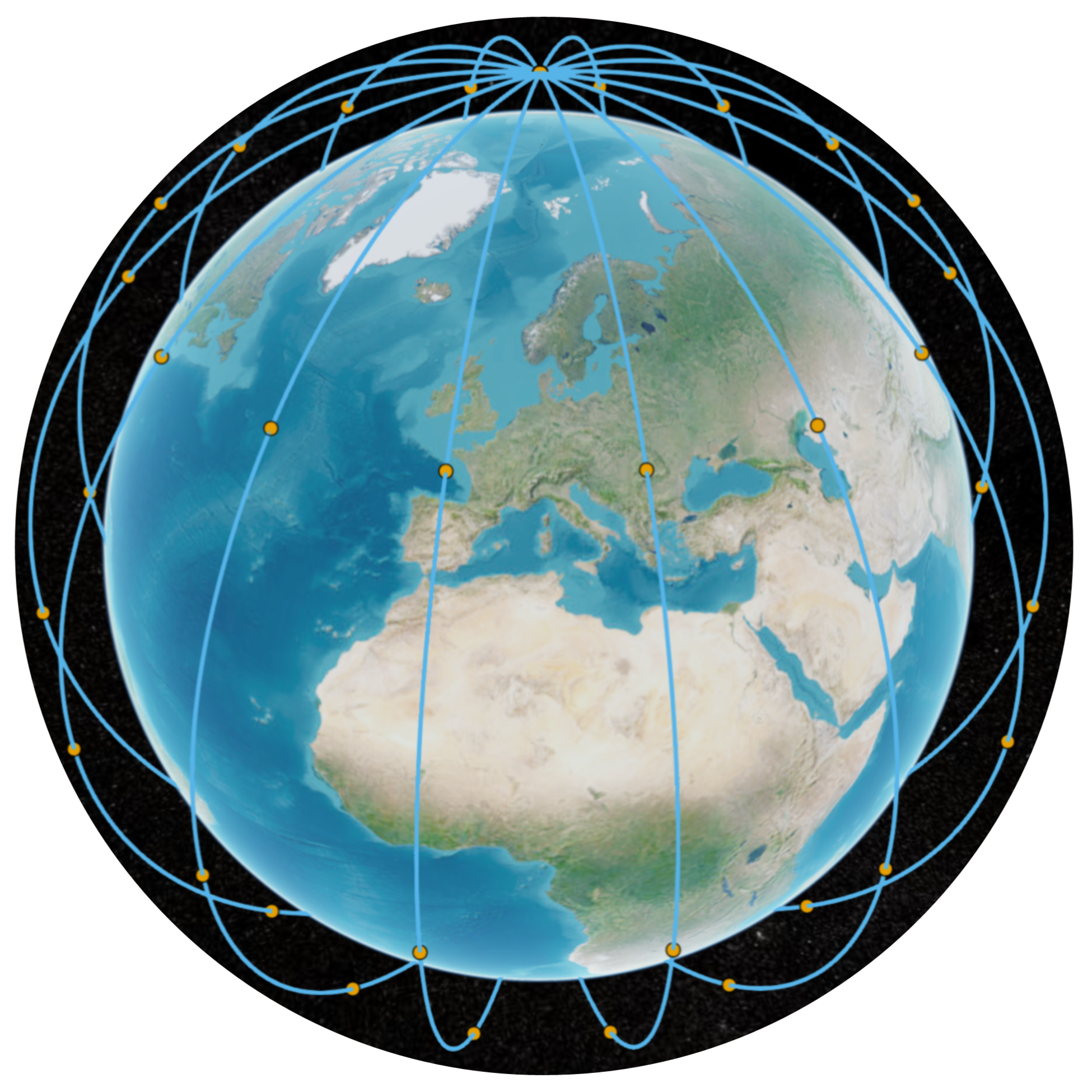}
	}
    \caption{(a) Walker Delta ($60\degree:64/8/0$) constellation and (b) Walker Star ($90\degree:64/8/0$) constellation. The altitude of the orbital shell has been set at 1,000 km.} 
	\label{fig:delta_vs_star}
\end{figure*}

IRIS\textsuperscript{2} represents an initiative from the European Union currently under development, aimed at deploying a secure and resilient multi-orbit satellite communication infrastructure \cite{iris_eur_lex}. Unlike other constellations listed in Table \ref{tab:comparacion_tabla}, IRIS\textsuperscript{2} adopts a hybrid architecture combining LEO and MEO satellites for a planned total count of approximately 290. Spectrum access and international coordination for IRIS\textsuperscript{2} are being pursued through non-geostationary satellite network filings submitted to the International Telecommunication Union (ITU) radiocommunication bureau \cite{iris_itu}. The LEO segment is expected to operate at altitudes around 1,200 km, while the MEO component would reach altitudes of up to 8,000 km, enabling a balance between coverage continuity, resilience and system redundancy. Operating mainly in the Ku, Ka and K bands, IRIS\textsuperscript{2} is primarily intended to support secure governmental communications, critical infrastructure protection and commercial broadband services within Europe.

Following a qualitative overview of existing LEO constellations and those planned for the near future, the following sections explore aspects of satellite orbit design applicable to the principles of the aforementioned constellations, as well as new satellite constellation designs. In particular, note that the described solutions have a global focus or target mid- and low-latitudes where population density is higher compared to regions of the North Atlantic. This means that constellations with a specific focus on this region remain largely unexplored. For these reasons, the analysis in the remainder of this work focuses on coverage at high latitudes, and more specifically on the North Atlantic.

\section{LEO Constellation Coverage}

For a given satellite constellation, the positions and dynamics of the satellites are deterministic with respect to specific positions on the ground, where the ground stations and users are located. Thus, with the geometric knowledge of the parameters of a constellation, it is possible to determine the likelihood of establishing visibility between a constellation and a terrestrial location. Therefore, a Walker-type orbital shell can be mathematically defined based on four parameters using the notation $i:T/P/F$, where $i$ is the orbital inclination of the constellation, $T$ is the number of satellites deployed in the orbital shell, $P$ is the total number of defined orbital planes, and $F$ is the phasing parameter that defines the relative spacing between satellites in adjacent planes. Intuitively, F indicates how many satellite positions the pattern shifts when moving from one orbital plane to the next, and defines the shift in argument of latitude between satellites in adjacent orbital planes, equal to $F \cdot 360\degree / P$ (Walker Delta) or $F \cdot 180\degree / P$ (Walker Star). The main difference between Walker Delta and Walker Star constellations lies in how the orbital planes are distributed in the Right Ascension of the Ascending Node (RAAN)~\cite{RAAN}. While in Walker Delta they are distributed over 360$\degree$, in Walker Star they are distributed over 180$\degree$. Fig.~\ref{fig:delta_vs_star} shows an example of a constellation deployment using Walker Delta geometry with $60\degree$ inclination ($60\degree:64/8/0$), compared to Walker Star geometry with $90\degree$ inclination ($90\degree:64/8/0$). As  it can be observed by comparing the two geometries, the main advantage of Walker Star lies in the high density of orbits converging toward the poles, which maximizes coverage in polar regions. However, this same geometry causes the distance between orbits to increase as we move toward latitudes near the equator. This makes it difficult to establish inter-satellite links at low latitudes and increases the likelihood of coverage gaps in certain equatorial regions for constellations with a small number of satellites. In contrast, Walker Delta geometries tend to be deployed at latitudes far from the poles, leaving these regions uncovered as seen in the polar gap for Fig.~\ref{fig:delta_vs_star}(a), in exchange for maximizing density in lower orbits. In the example shown in Fig.~\ref{fig:delta_vs_star}(a), the orbital density is maximized around the inclination defined by the geometry, i.e., 60\degree.  As with Walker Star orbits, coverage decreases for regions near the equator in the case of constellations with few satellites due to the Earth’s larger equatorial diameter. The probability of coverage based on constellation geometry is analyzed and quantified in more detail in Section~IV.

\begin{figure}[!t]
\centering
\includegraphics[width=0.8\linewidth]{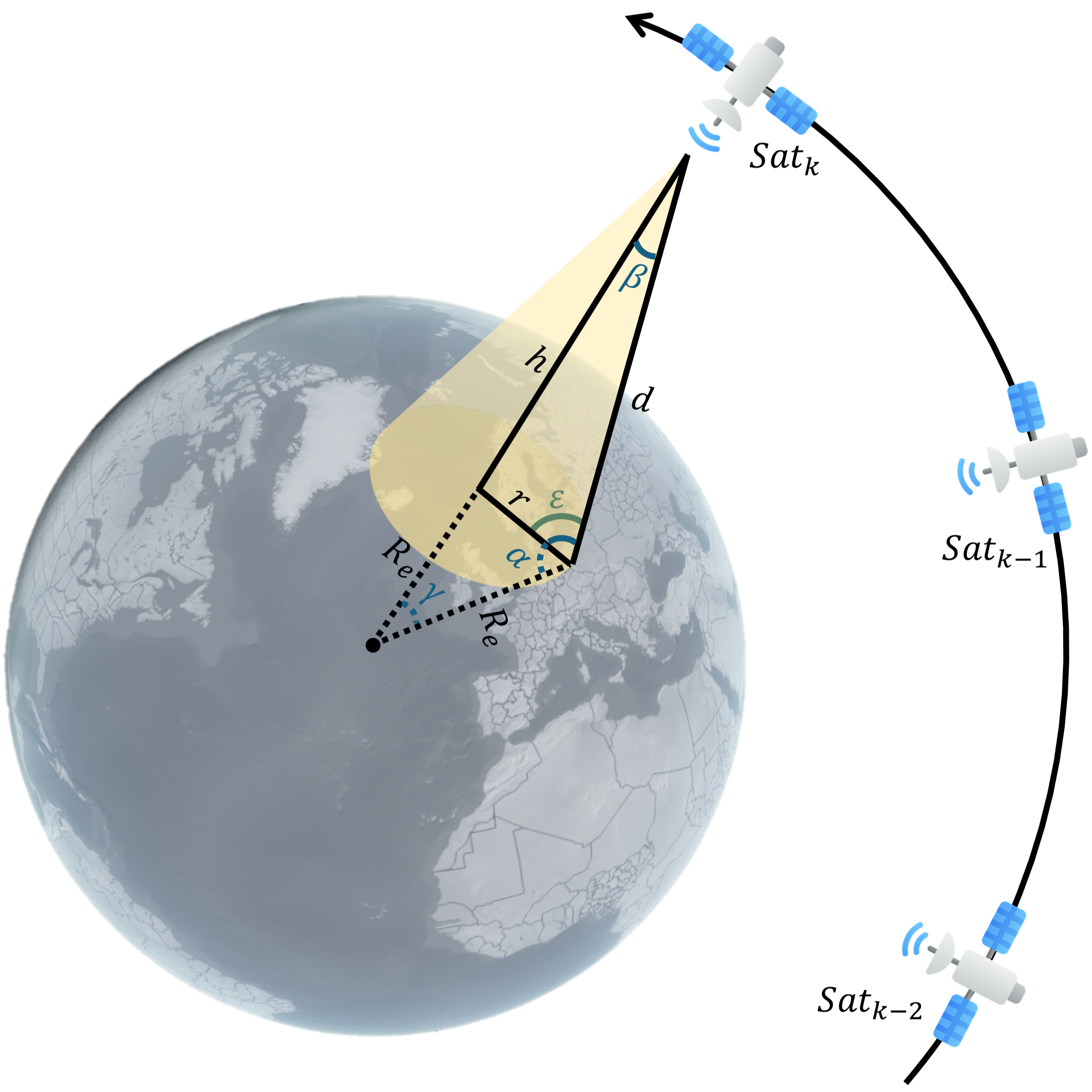}
\caption{Projection of the satellite's visibility relative to the Earth's surface for a minimum elevation angle $\epsilon$, and the triangle formed between (i) the center of the Earth, (ii) the satellite, and (iii) a user on ground located at the edge of the visibility zone. The objects and distances are not to scale and are for illustrative purposes only.}
\label{fig:esquema_visibilidad}
\end{figure}

\begin{figure}[!b]
	\centering
	\subfigure[]{\includegraphics[width=1\columnwidth]{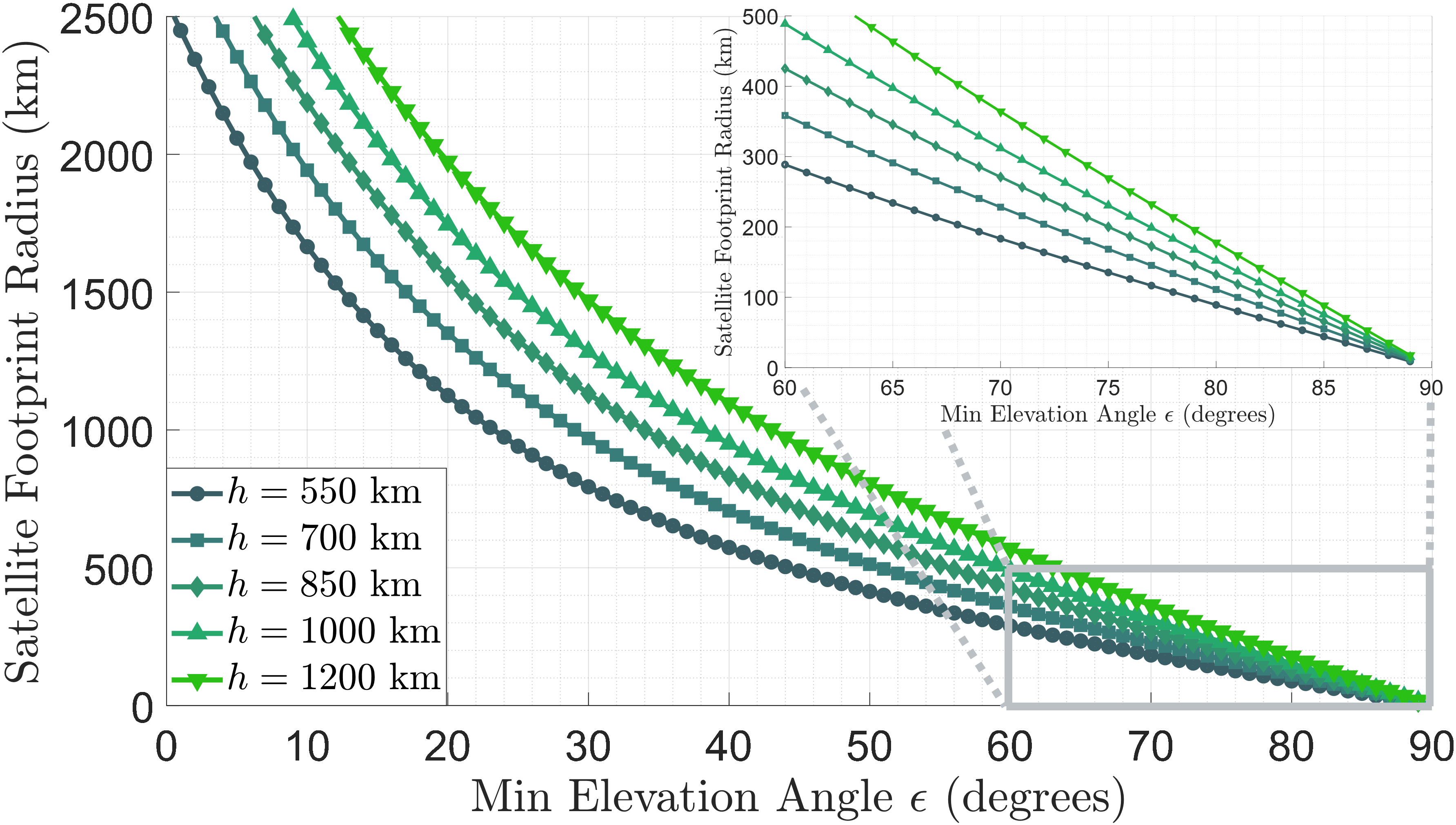}
	} 
    \subfigure[]{\includegraphics[width=1\columnwidth]{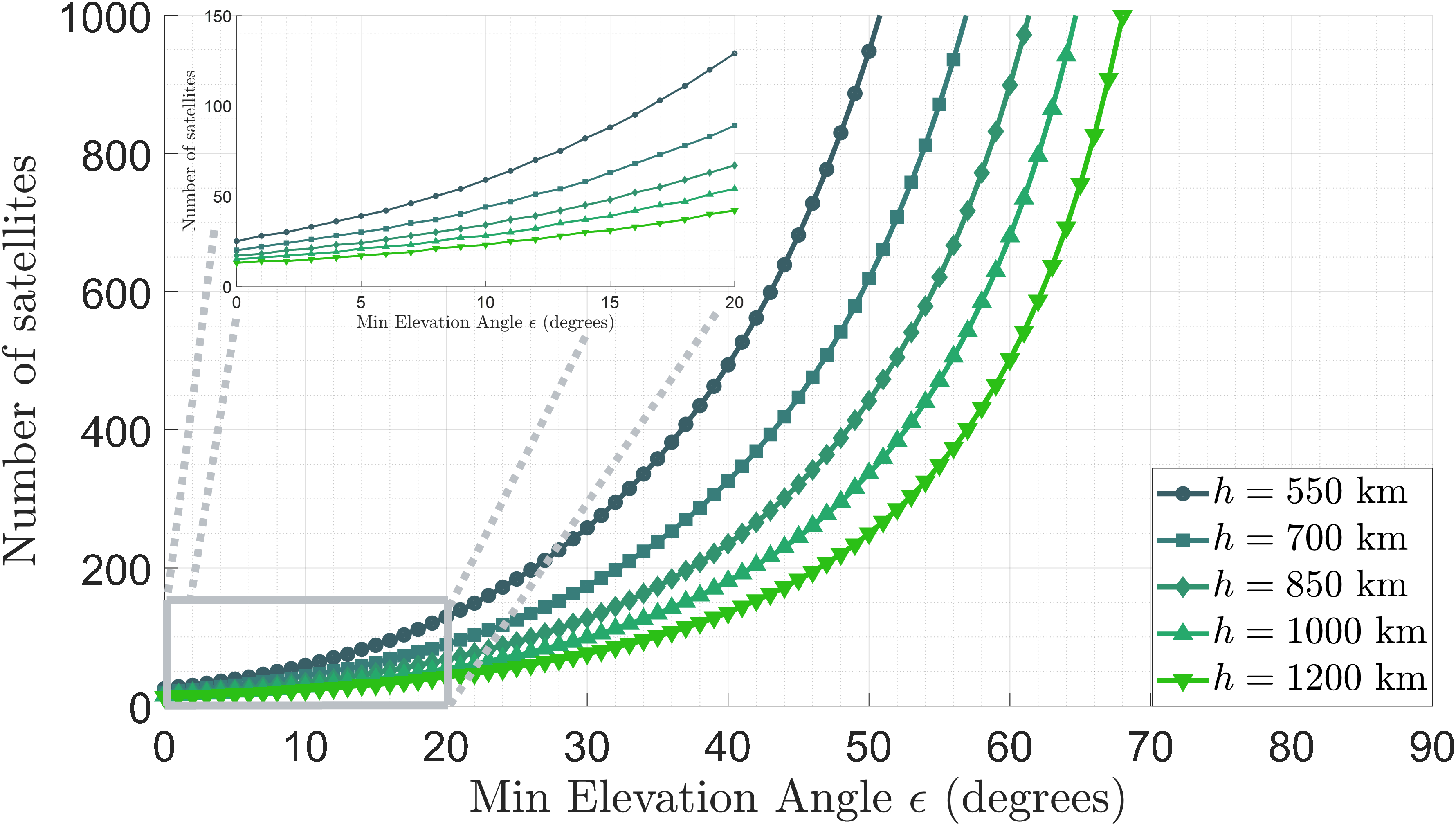}
	}
    \caption{(a) Coverage area radius of a satellite footprint and (b) lower bound of the number of satellites required to establish visibility in an area equivalent to the Earth's surface for several minimum elevation angles $\epsilon$ between user and satellite, and several altitudes $h$ of constellations in LEO orbits.} 
	\label{fig:elevation_vs_radioandnumber}
\end{figure}

In addition to the previous parameters that define orbital characteristics, the altitude of the constellation $h$ and the minimum elevation angle between the user on the ground and the satellite $\epsilon$ determine the visibility conditions between the actors involved. Geometrically, these last two parameters define the projection of a given satellite's coverage on Earth. Let us assume a user located at the edge of the visibility cell with radius $r$ projected by a given satellite. Through the triangle formed between the center of the Earth, the satellite, and the user (see Fig.~\ref{fig:esquema_visibilidad}), the law of sines can be applied as follows:

\begin{equation}
\frac{R_e+h}{\sin (\alpha)}=\frac{R_e}{\sin (\beta)},
\end{equation}

\noindent where $R_e$ is the radius of the Earth, which can be approximated as 6,378 km. Solving $\beta$ from (1), and given that $\alpha = \pi/2 + \epsilon$, since the Earth's radius is perpendicular to the local horizon, we can calculate the radius of the projection on Earth as:

\begin{equation}
r = \gamma R_e = (\pi - \alpha - \beta)R_e = \left(\arccos \left(\frac{R_e \cdot \cos \left(\varepsilon\right)}{R_e+h}\right)-\varepsilon\right) R_e.
\end{equation}

\begin{figure}[!t]
\centering
\includegraphics[width=1\linewidth]{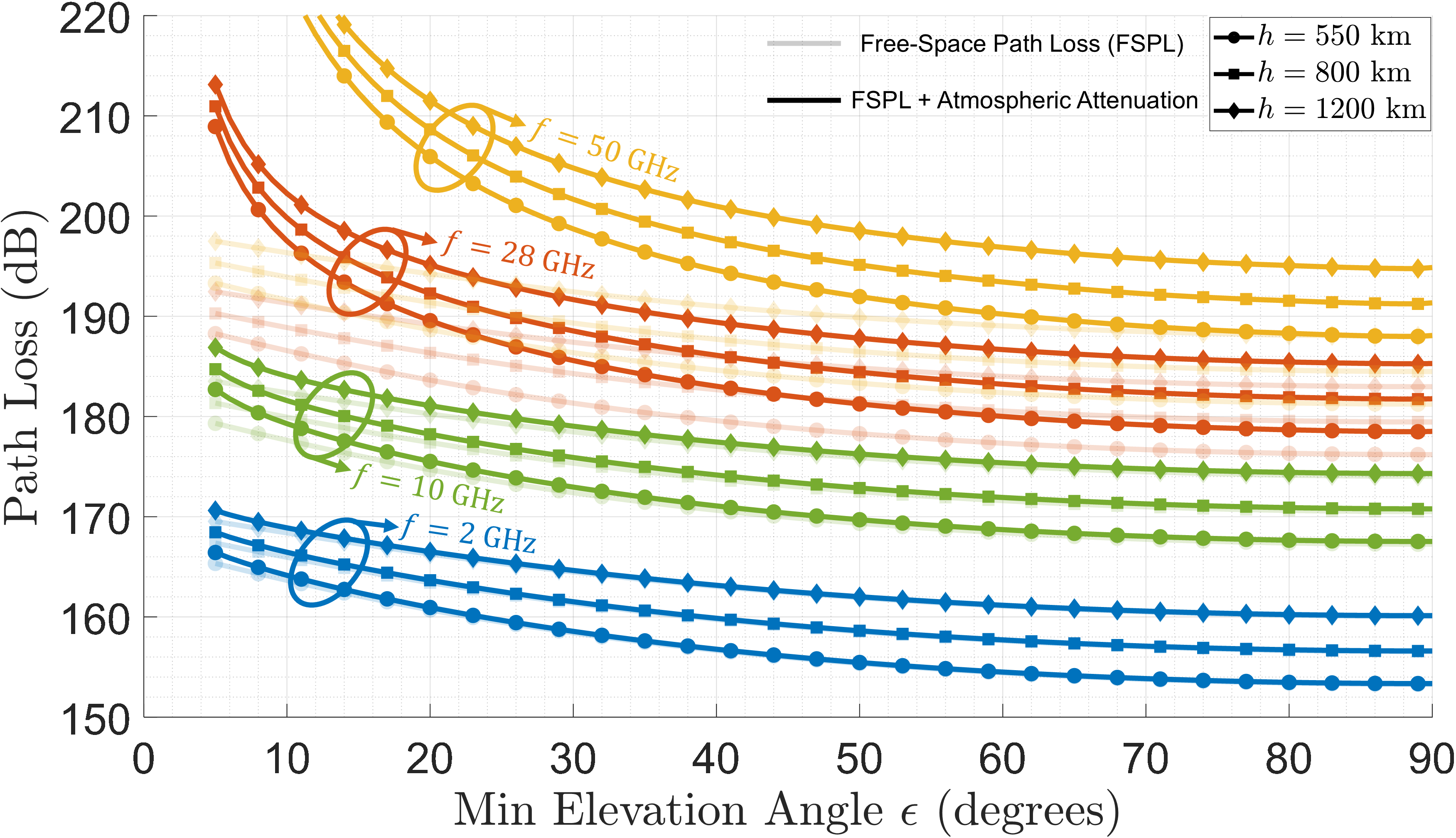}
\caption{Path loss for an Earth-space communications system deployed in LEO orbits, considering the attenuation due solely to free-space path loss, as well as the combined free-space path loss and atmospheric attenuation for four frequencies $f$ (2~GHz, 10~GHz, 28~GHz and 50~GHz), three altitudes $h$ (550~km, 800~km and 1200~km) and several elevation angles~$\epsilon$.}
\label{fig:PL_vs_elevation}
\end{figure}

Based on (2), and given the exclusive dependence of the radius of visibility on $h$ and $\epsilon$, a parametric sweep can be performed to determine the expected radius of the coverage area provided by a given satellite. Fig.~\ref{fig:elevation_vs_radioandnumber}(a) shows this coverage area for minimum elevation angles from $0\degree$ to $90\degree$, and with altitudes $h$ corresponding to different LEO orbits from 550~km to 1200~km. As expected, it can be seen that decreasing the altitude of the constellation decreases the effective coverage area of a satellite. For example, for an $\epsilon = 30\degree$, the coverage radius of the visibility area decreases from 1470 km to 793 km when the altitude $h$ decreases from 1200 km to 550 km. Similarly, increasing the restriction around the minimum elevation angle $\epsilon$ causes convergence towards a null radius as this parameter tend towards $\epsilon = 90\degree$. For an area of the Earth equal to $A_e=510,072,000 \textrm{ km}^2$, the number of satellites required to provide global coverage, i.e., visibility with a satellite from any location on the Earth's surface, can be approximated as $\left\lceil A_e/(\pi r^2)\right\rceil$. Note that this approximation is a theoretical lower bound, since the circular footprint of each satellite cannot cover the Earth's surface homogeneously, so in practice the number of satellites required is slightly higher. Fig.~\ref{fig:elevation_vs_radioandnumber}(b) shows the minimum number of satellites required for providing coverage to an area equivalent to the Earth's surface as a function of $\epsilon$ and $h$. For constraints of $\epsilon$ close to $0\degree$, it is theoretically possible to establish global visibility with even fewer than 50 satellites in total. However, this number increases rapidly for $\epsilon > 30\degree$ due to the drastic reduction in footprint size.

Based on the assumptions of the previous analysis, it seems reasonable to infer that a global coverage deployment is possible given a small number of satellites with a low $\epsilon$. However, in practice, this deployment is not feasible for a number of reasons. Firstly, the antennas available on both the user terminal and the satellite are not designed for covering elevation angles close to the horizon. This is because, although these components are made up of hundreds and thousands of elements with beamforming capabilities, the arrays exhibit limited beamforming performance at large deviations from the specular direction. For example, the \textit{Standard} Starlink~\cite{starlink_standard_dish} user terminal has a field of view of $110\degree$, while the \textit{Performance} Starlink~\cite{starlink_performance_dish} user terminal has a field of view of $140\degree$. Equivalently, these characteristics geometrically limit the minimum elevation angle to $\epsilon = 35\degree$ and $\epsilon = 20\degree$, respectively. 

In addition, the link budget is compromised as the satellite moves away from the zenith from the user's point of view. Apart from the loss of antenna gain itself, the effective distance between the user terminal and the satellite increases, which increases free-space attenuation, as well as losses due to atmospheric gases, and ionospheric and tropospheric scintillation~\cite{ionospheric_attenuation}. Therefore, the channel path loss due to propagation effects can be expressed as:

\begin{equation}
PL_t(\mathrm{dB}) = PL_{friis}(\mathrm{dB})+PL_g(\mathrm{dB})+PL_s(\mathrm{dB}),
\end{equation}

\noindent where $PL_t$ is the total path loss, $PL_{friis}$ stands for the free-space path loss, $PL_g$ is the path loss due to atmospheric gases, and $PL_s$ accounts for the ionospheric and tropospheric
scintillation. Note that the previous formula does not include any term for blocking or outdoor-to-indoor losses, since it assumes the ground-based antenna has LoS towards the satellite in space. Free-space path loss is calculated by taking into account the increase in distance between the user and the satellite as the elevation angle decreases, considering a path loss exponent of two, as specified by the Friis formula~\cite{Friis}. Therefore, $PL_{friis}$ is calculated as:

\begin{equation}
PL_{friis}(\mathrm{dB}) = 20 \log _{10}\left(\frac{c}{4 \pi d f}\right),
\end{equation}

\noindent where $c$ is the speed of light, $d$ is the distance between the satellite and the user terminal on the ground, and $f$ is the operating frequency.

On the other hand, $PL_g(\mathrm{dB})$ and $PL_s(\mathrm{dB})$ are modeled by: (i) the altitude $h$ and the minimum elevation angle $\epsilon$, since they define the link distance $d$, and therefore determine the path of electromagnetic waves through the atmosphere, (ii) the operating frequency $f$, since attenuation by atmospheric gases such as oxygen or water vapor depends heavily on frequency, and (iii) the geographic location of the user terminal, due to statistics
such as precipitation used in the calculation of these parameters depend on geographic coordinates. The calculation of both parameters is standardized by the International Telecommunication Union and it is defined in ITU-R Recommendations ITU-R P.618~\cite{P618}, ITU-R P.676~\cite{P676}, ITU-R P.834~\cite{P834}, and ITU-R P.531~\cite{P531}. Fig.~\ref{fig:PL_vs_elevation} shows the path loss considering only free-space path loss (FSPL), as well as the total losses, which also consider atmospheric attenuation, for four frequencies $f$ (2~GHz, 10~GHz, 28~GHz and 50~GHz) and three altitudes within the LEO orbit range.  For atmospheric attenuation calculations, the ground station is considered at 57.0138$\degree$ N, 9.9871$\degree$ E, thus examining the annual statistics of this location when modeling the characteristics of precipitation, as well as attenuation caused by clouds, fog, and water vapour~\cite{P836, P837, P840}. Focusing on the results, for $h = 800$ km and $\epsilon = 90\degree$, the FSPL is 157~dB, 171~dB, 179~dB, and 184~dB for frequencies of 2~GHz, 10~GHz, 28~GHz, and 50~GHz, respectively, indicating very high losses associated with the large range. These losses are even more pronounced for $\epsilon < 90\degree$ since the satellite is no longer at the zenith, and its distance with respect to the ground station increases as $\epsilon$ decreases. For instance, for $h = 800$ km and $\epsilon = 10\degree$, the FSPL increases to 166~dB, 180 dB, 189 dB, and 193 dB, representing a $\sim$10~dB increase in losses that must be accounted for as a margin in the link budget. With regard to atmospheric attenuation, it can be observed that the results are highly frequency-dependent. For $f = 2$ GHz, the differences relative to FSPL are practically negligible except for $\epsilon < 15\degree$, where deviations from FSPL of around 0.5–1~dB are observed. For $f = 10$ GHz, this atmospheric attenuation remains practically negligible for high values of $\epsilon$. However, the attenuation compared to FSPL begins to differ significantly for $\epsilon < 30\degree$, with differences of up to 3 dB observed for the smallest elevation angles. This is explained by the fact that these elevation angles involve longer propagation paths through the troposphere and ionosphere, as well as layers of clouds and rain, making the electromagnetic waves more prone to the losses described above. In the case of $f = 28$ GHz and $f = 50$ GHz, it is observed that even for $\epsilon = 90\degree$, there is already a 2 dB (28 GHz) and 7~dB (50 GHz) difference between FSPL and total losses, indicating the increased impact of atmospheric losses as frequency increases due to the shorter wavelength of the signal. This effect accumulates more rapidly compared to lower frequencies (10~GHz and 2~GHz), with losses of 5~dB (28~GHz) and 14~dB (50~GHz) for $\epsilon = 25\degree$, and 10 dB (28 GHz) and 32 dB (50 GHz) for $\epsilon = 10\degree$. Note the sharp increase in atmospheric attenuation losses at 50 GHz compared to lower bands due to the oxygen absorption band at 60 GHz. Despite the high losses, operators such as SpaceX are obtaining licenses to operate in this V-band due to the increased bandwidth available~\cite{vband_starlink}, with the frequencies shown in Fig.~\ref{fig:PL_vs_elevation} being representative of the frequency bands used by satellite operators (see Table~\ref{tab:comparacion_tabla}). Note that the results shown apply exclusively to the path loss of the propagation channel and are therefore independent of the specific hardware of any satellite constellation. In order to consider the complete link budget, we must also take into account, in addition to the total path loss $PL_t$, the transmitted $P_{t_{Sat./UT}}$ and received $P_{r_{UT/Sat.}}$ powers at both the user terminal and the satellite, as well as the antenna gains at the user terminal $G_{UT}$ and the satellite $G_{Sat.}$. Thus, the relation among all the variables is given as $P_{r_{UT/Sat.}}(\mathrm{dBW}) = P_{t_{Sat./UT}}(\mathrm{dBW}) + G_{Sat.}(\mathrm{dB}) + G_{UT}(\mathrm{dB}) - PL_t(\mathrm{dB})$. Unfortunately, satellite operators generally use proprietary solutions for which no specific information regarding transmission power and antenna gain is made public. However, it is possible to obtain information on Effective Isotropic Radiated Power (EIRP) from Federal Communications Commission (FCC) applications, which can be used to make rough estimations of the link budget~\cite{fcc_starlink_eirp}.

\begin{figure}
\centering
\includegraphics[width=1\linewidth]{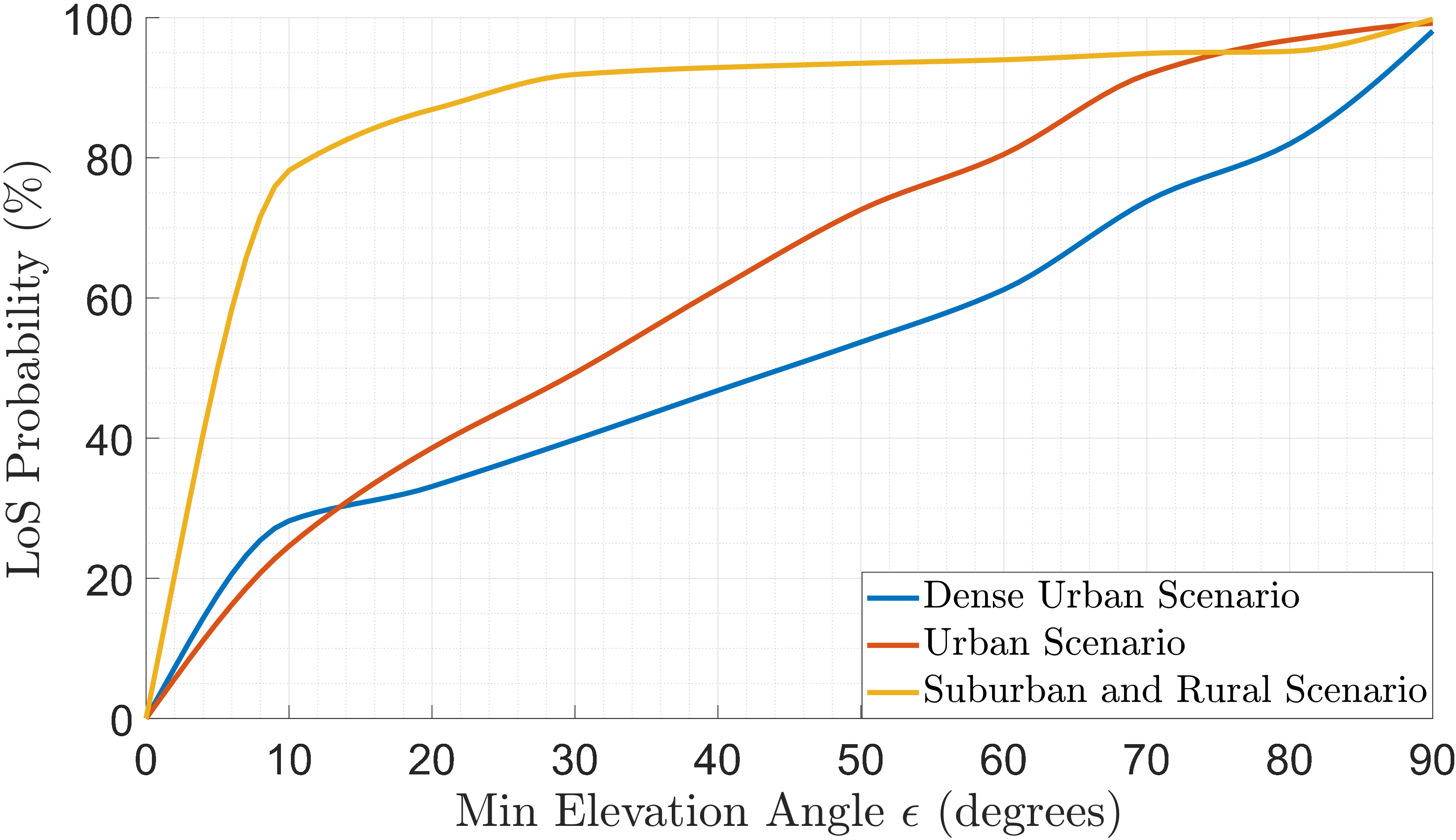}
\caption{Line-of-Sight probability for an Earth-space communications system deployed in LEO orbits for (i) dense urban, (ii) urban, and (iii) suburban scenarios given several elevation angles $\epsilon$.}
\label{fig:LoS_probability}
\end{figure}

Finally, the probability of establishing Line-of-Sight (LoS) is also critical. Non-terrestrial communications systems require LoS conditions, as blockage losses in NLoS at operating frequencies make it impossible to establish a communications link. Fig.~\ref{fig:LoS_probability} shows the probability of LoS in three types of scenarios: (i) dense urban, (ii) urban, and (iii) suburban and rural scenarios according to 3GPP technical report TR38.811~\cite{3gpp_tr38811}. For elevation angles close to the horizon, e.g., $\epsilon = 10 \degree$, the probability of LoS is 78.2\% in rural areas. However, this value drops to 28\% and 24\% in dense urban and urban areas, respectively. To obtain a LoS probability of 50\%, an $\epsilon$ of at least $45\degree$ is required in dense urban areas, and $30\degree$ in urban areas. For a $90\%$ LoS probability, these angles increase to $85\degree$ and $70\degree$, respectively.

Thus, previous analysis highlights the challenge of establishing visibility with satellites far from the zenith. In practice, each satellite operator establishes a minimum elevation mask for connecting a satellite to a user terminal~\cite{satellite-comparison, min_elevation_oneweb, min_elevation_iridium}. The previous challenges in terms of path loss and minimum elevation angle require the development of an efficient and satellite constellation design that can deploy a small number of satellites while providing the widest possible ground coverage. An efficient design minimizes the already high cost of deploying satellites in low Earth orbits, while also positively impacting other aspects, such as reducing space debris once the satellite’s useful life has ended, or reducing the risk of collisions in the event of a crowded LEO space region. These aspects of the design of the orbital parameters for a LEO constellation are covered in Section~IV.\\

\section{Design of LEO Constellations for North Atlantic Coverage}

As outlined in Sections I and II, most commercial LEO constellations fall into two categories: (i) those designed with inclinations of $i < 55\degree$, which aim to optimize the number of satellites required to provide effective coverage of the Earth’s most densely populated areas, and (ii) those designed with polar orbits, i.e. $i \approx 90\degree$, which aim to provide global visibility, albeit at the cost of over-coverage at the poles. Given the previous scenarios, low-inclination constellations provide poor coverage at high latitudes, e.g., the North Atlantic and Arctic regions, whilst polar constellations over-cover non-priority areas if the area of interest lies at inclinations with latitudes below 90$\degree$. Thus, there is a critical region between the two designs that is not optimized for either scenario. This section aims to analyze, through a parametric analysis in which each parameter is varied independently, some of the key metrics used to determine the connectivity capacity of a ground-based user terminal to satellites in orbit, considering latitudes between $55\degree$ N and $90\degree$ N, and parameters related to the constellation’s geometry, as described in Section~III.

To carry out the analysis, the positions and dynamics of the satellites are simulated using the Simplified Perturbation Model (SGP4), and the orbital characteristics of each satellite are defined using two-line element sets (TLEs). SGP4 is an orbit propagator designed to calculate the position and velocity of a satellite in space based on perturbations and the influence of gravitational fields from both the Earth and other celestial bodies such as the Moon and the Sun. The reason we have opted for simulations using SGP4 rather than analytical models is that we aim to take into account (i) the specific dynamics and parameters of a Walker Delta-type constellation, that is, its $i:T/P/F$ geometry plus the altitude $h$ and the minimum elevation angle $\epsilon$, as well as (ii) the specific position in terms of latitude and longitude of a ground-based user terminal. To the best of the authors' knowledge, analytical expressions for visibility and coverage available in the literature generally rely on simplifying assumptions regarding the spatial distribution of satellites (e.g., uniform distributions)~\cite{coverage_dense_networks}. Although these approaches take into account the number of satellites and the footprint of each satellite, i.e., $h$ and $\epsilon$, the assumptions on the spatial distribution neglect the deterministic orbital geometry imposed by the Walker Delta parameters, the resulting temporal evolution of the constellation, and the location of the user terminal. Therefore, those analytical formulations cannot be directly applied to the scenario under study considered in this work. 

The simulation analysis has been conducted by simulating satellite dynamics over 5-day periods with 10-second time steps, i.e., granularity between satellite positions. Given the speed of the satellites and the altitudes in LEO orbits, orbital periods range from approximately 94~minutes (500~km) to 109~minutes (1200~km), resulting in the analysis of between 66 and 77 complete orbits in 5-day simulations. For the sake of clarity, multiple simulation durations and time step values have been evaluated, confirming that the selected parameters ensure the convergence of the different metrics analyzed throughout this section while providing a suitable trade-off in terms of overall simulation time. In particular, simulations have been conducted with durations ranging from 1 to 14 days to verify the convergence of the various metrics studied. Analysis of the results revealed a maximum variation of 0.2\% in the considered metrics across simulation durations of 5 to 14 days, indicating the convergence of the results for a 5-day simulation period, thereby seeking a trade-off between accuracy and simulation time. Similarly, multiple simulation time steps ranging from 3 s to 180 s have been studied. It is observed that for time steps larger than 30 s, the simulation begins to omit short visibility events occurring within this time window, leading to increasingly different results compared with the converged values obtained with smaller time steps. However, below this value, the variation of the different metrics converges, with differences of less than 0.1\% observed when considering time steps of 3 s and 10 s. For the reasons stated above, the time step is therefore set to 10 s. Finally, given the finite nature of the observation period, the effect of the constellation’s initial phase on the simulations has been studied. To this end, several initial phases have been considered and across them, the maximum variation among the considered metrics is 0.3\%, indicating that the selected simulation duration provides sufficiently stable results with respect to the initial constellation phase.

Among the most relevant metrics for defining service capacity are (i) constellation availability, defined as the probability that a ground station will have LoS condition to at least one satellite in the constellation, and (ii) revisit time, defined as the time it takes for a ground station to recover satellite visibility in the constellation if it has been lost to a previous satellite.

\sloppy Focusing on the simulated constellation availability, given a ground station at specific terrestrial coordinates $\mathbf{r}_g(Lat., Lon.)$, and a set of satellites located at time $t$ at positions \mbox{$\mathbf{r}_k(t,i,T,P,F,h)$} with $k = 1,...,T$, the metric that constitutes simulated visibility between $\mathbf{r}_g$ and $\mathbf{r}_k$ is calculated as

\begin{equation}
v_k(t,\epsilon,\mathrm{Lat.},\mathrm{Lon.})=
\begin{cases}
1 &
\parbox[t]{0.25\textwidth}{
if satellite $k$ is visible from $(\mathrm{Lat.},\mathrm{Lon.})$
with elevation mask $\epsilon$
} \\[1ex]
0 & \text{otherwise}
\end{cases}.
\end{equation}

\noindent Therefore, the total number of visible satellites is

\begin{equation}
N_{\mathrm{sat}}(t, \epsilon, Lat., Lon.)=\sum_{k=1}^T v_k(t, \epsilon, Lat., Lon.), 
\end{equation}

\noindent and the probability of establishing visibility with at least one satellite along the simulated scenario is

\begin{equation}
\begin{split}
P\left(
N_{\mathrm{sat}} \geq 1
\mid
\mathrm{Lat.}, \mathrm{Lon.}, i, T, P, F, h
\right)
= {} & \\
&\hspace{-2cm} \mathbb{P}\left(
\sum_{k=1}^{T}
v_k(t,\epsilon,\mathrm{Lat.},\mathrm{Lon.})
\geq 1
\right).
\end{split}
\end{equation}

\noindent Note that this probability of visibility refers to the geometric visibility, which is the minimum condition required to eventually establish a usable communication link. Establishing such a link will depend on the hardware in the specific system deployed (e.g., equipment sensitivity) and on physical channel conditions such as the path loss discussed in Section~III.

\sloppy Concerning revisit time, at a location $\mathbf{r}_g$ and time $t$, visibility $V(r_g,t)$ is achieved if $\sum_{k=1}^T v_k\left(\mathbf{r}_g, t\right) \geq 1$.  Therefore, the revisit time at event $n$ will be determined by the loss of visibility at time $t_{n,loss}=\left\{t: V\left(t^{-}\right)=1, V(t)=0\right\}$ and the restoration of visibility at time $t_{n,recover}=\left\{t: V\left(t^{-}\right)=0, V(t)=1\right\}$. This leads to the revisit time at event $n$ be given by $\tau_n =  t_{n,recover} - t_{n,loss}$. Finally, the set of events $n$ gives rise to the distribution of the revisit time $\tau$ for the designed constellation.

Given the preceding metrics, the following subsections provide a detailed analysis of the constellation’s behavior based on parameters such as the location of the user or ground station, or parameters intrinsic to the Walker Delta constellation, such as the elevation angle $\epsilon$ or the inclination $i$. Other constellation parameters, such as the number of satellites, number of orbits, phasing, and constellation altitude, have been set to achieve a trade-off in the analysis and the conclusions drawn from the study. In particular, the total number of satellites in the constellation has been set to $T = 64$ satellites. The reasoning behind this is to emulate the behavior of medium-sized constellations on the order of magnitude of solutions such as Iridium NEXT or GlobalStar (see Table~\ref{tab:comparacion_tabla}). Increasing the number of satellites, equivalent to that of mega-constellations such as Starlink or OneWeb, is no longer relevant, since such a large number of satellites in orbit implies continuous coverage practically independent of the satellites’ orbital configuration due to the redundancy of satellites in space. On the other hand, constellations with a very small number of satellites (e.g., T < $\sim$10 satellites) result in continuous visibility gaps regardless of the design of the orbital parameters, making it difficult to evaluate the suitability of different configurations due to the small number of satellites. Furthermore, $P = 8$ orbital planes has been set, thereby ensuring a homogeneous distribution of orbital planes and the number of satellites per plane. This prevents an excessive number of satellites from being in the same orbital plane, which could lead to overlap between satellites in the same orbit, while also leaving certain regions on Earth without any satellites orbiting nearby. The phasing has been set to $F = 3$ since $F \neq 0$ ensures that satellites in different orbital planes are longitudinally offset, avoiding persistent parallel alignment~\cite{phasing}. This results in improved global coverage uniformity. Finally, an altitude of $h = 1000$~km has been chosen as an intermediate trade-off within the spectrum of LEO altitudes.

\begin{figure*}[!b]
\centering
\subfigure[]
{\includegraphics[width=0.59\linewidth]{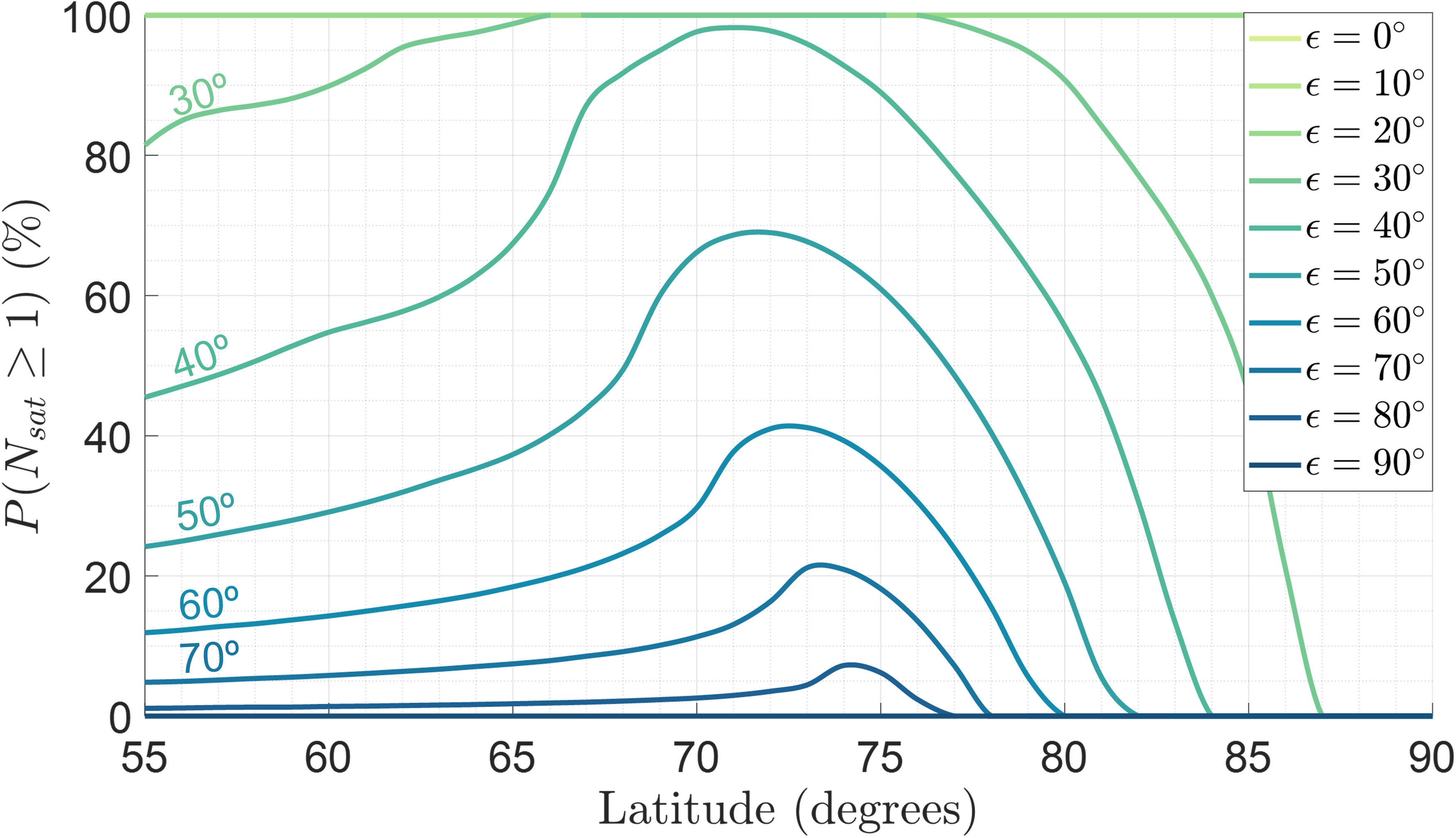}
}
\subfigure[]{\includegraphics[width=0.39\linewidth]{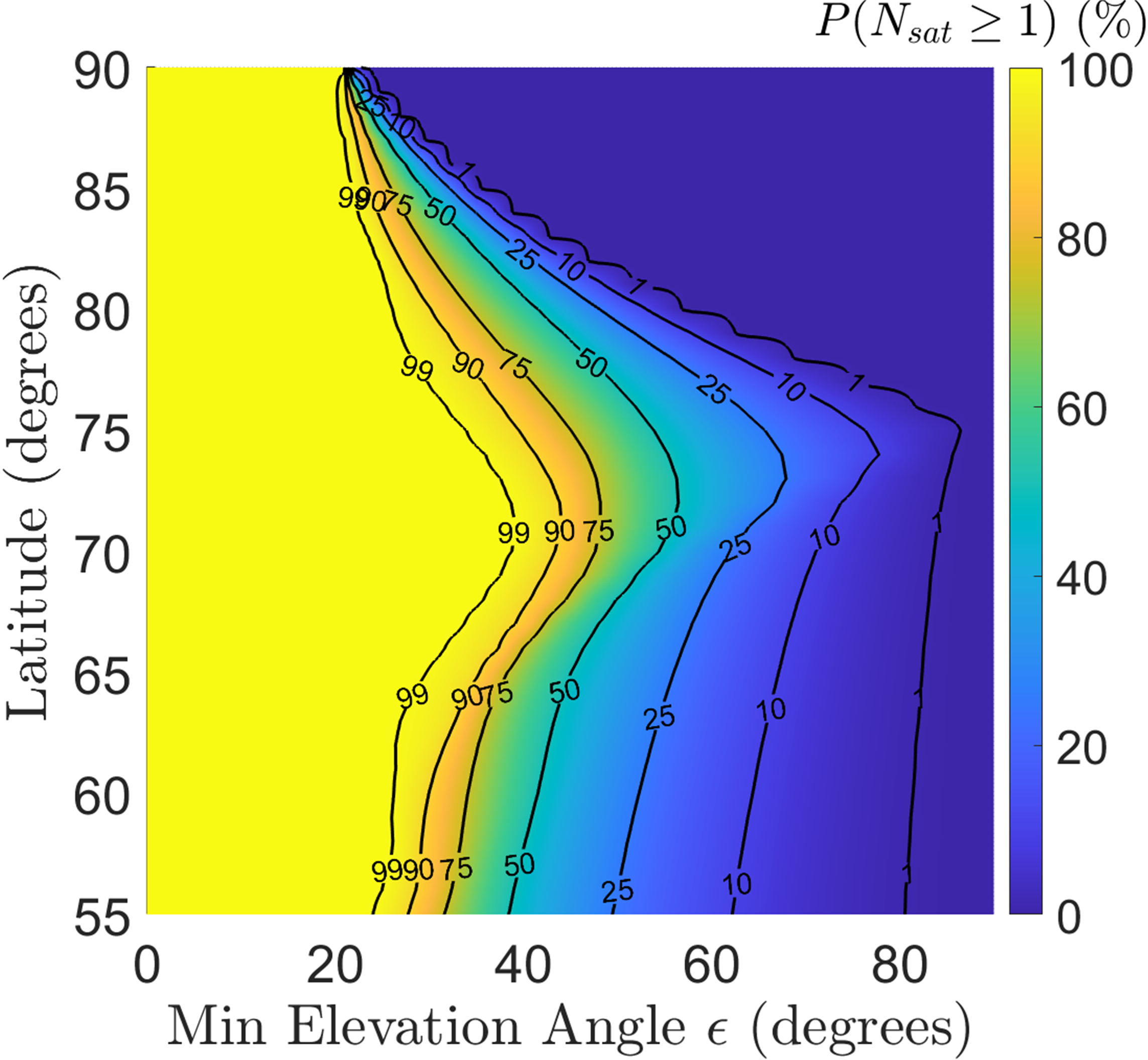}
}
\caption{(a) Probability of establishing visibility with at least one satellite in the LEO constellation for several latitudes and elevation angles $\epsilon$, and (b) isolines for $P(N_{sat} \geq 1)$ at 99\%, 90\%, 75\%, 50\%, 25\%, 10\%, and 1\%. Results for Walker Delta ($75\degree:64/8/3$) constellation with $h = 1000$ km.}
\label{fig:sweep_elevacion}
\end{figure*}

\begin{figure*}[!b]
	\centering
	\subfigure[{\footnotesize $\epsilon = 20\degree$}]{\includegraphics[width=0.32\linewidth]{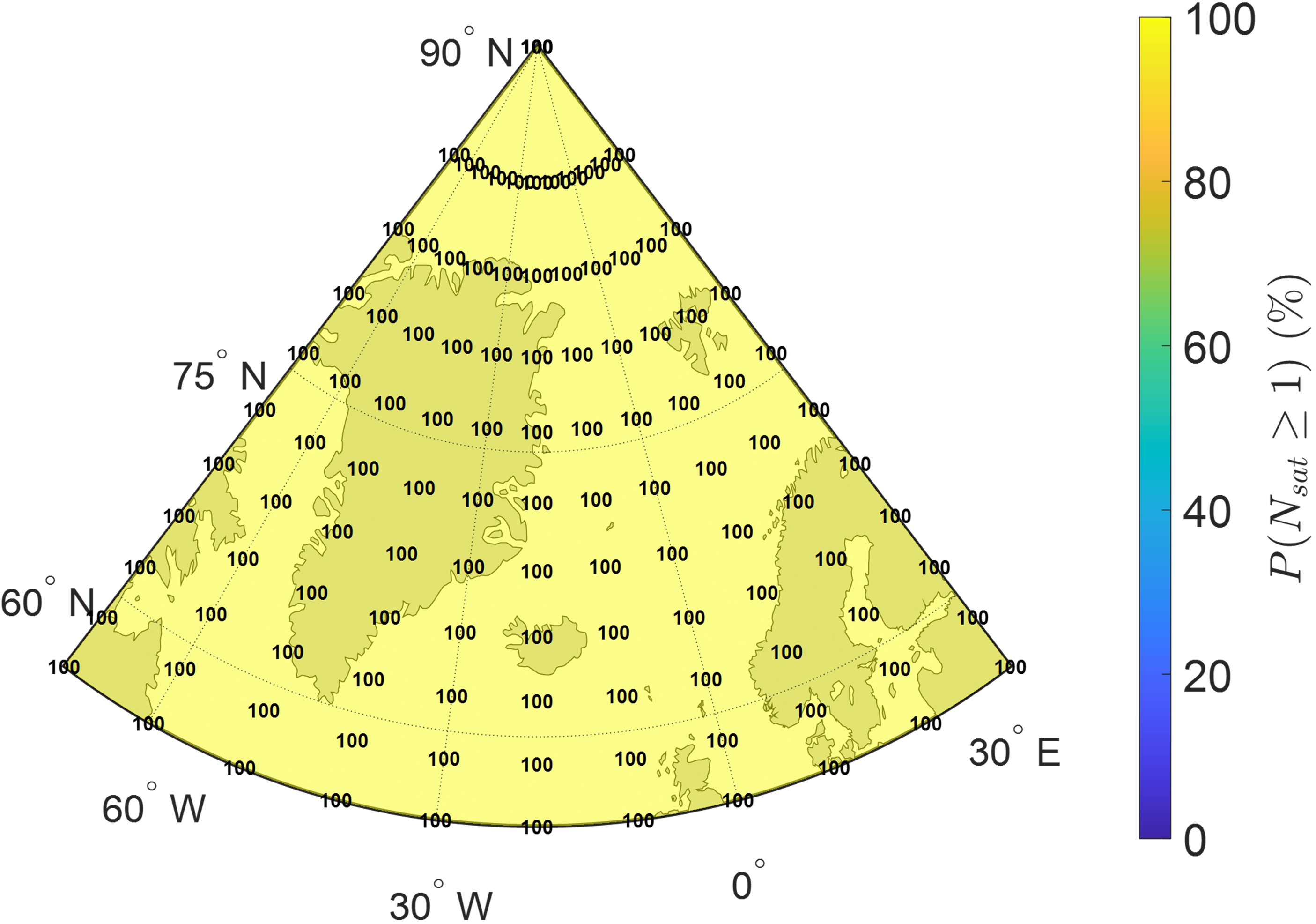}
	} 
    \subfigure[{\footnotesize $\epsilon = 40\degree$}]{\includegraphics[width=0.32\linewidth]{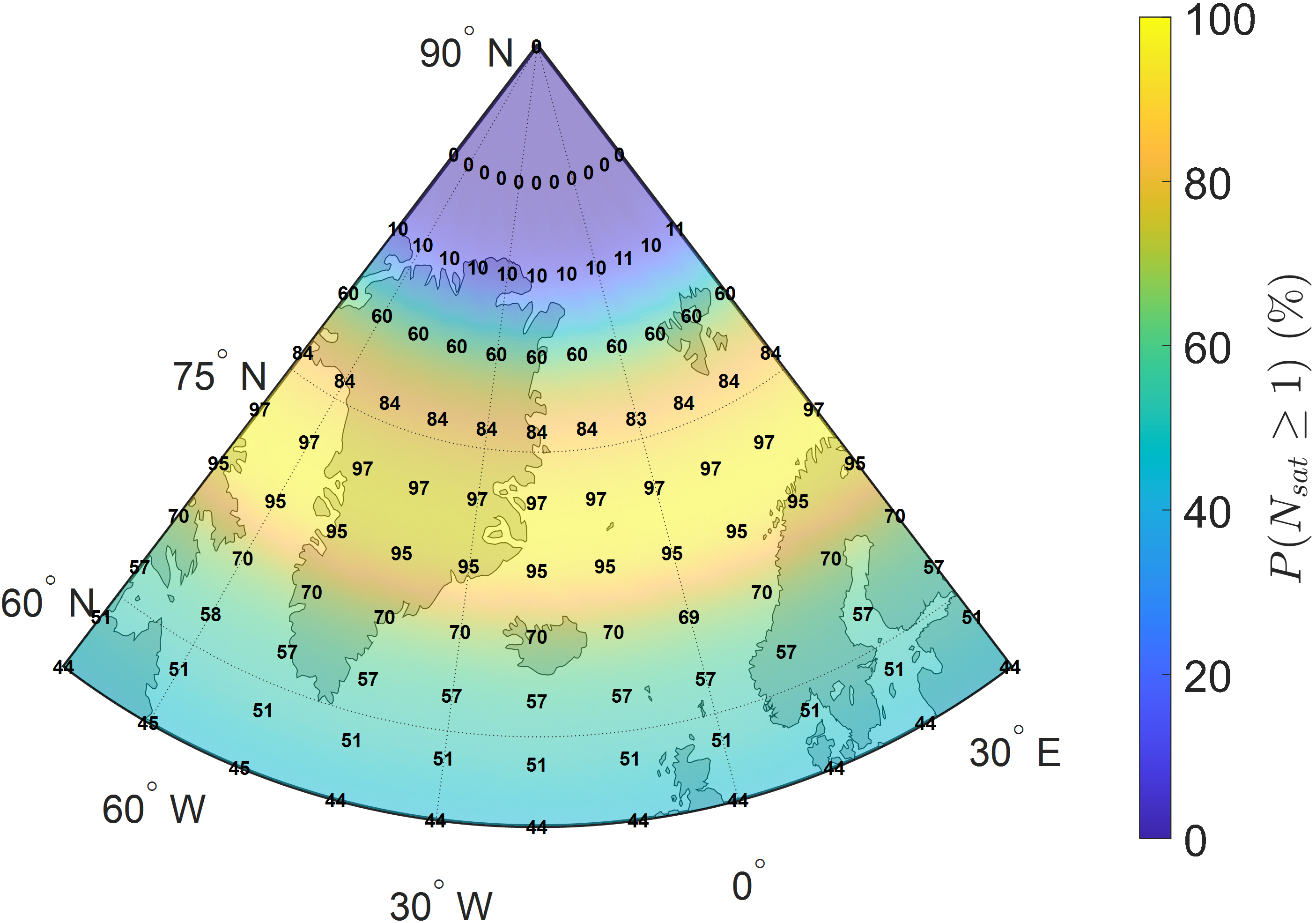}
	}
    \subfigure[{\footnotesize $\epsilon = 60\degree$}]{\includegraphics[width=0.32\linewidth]{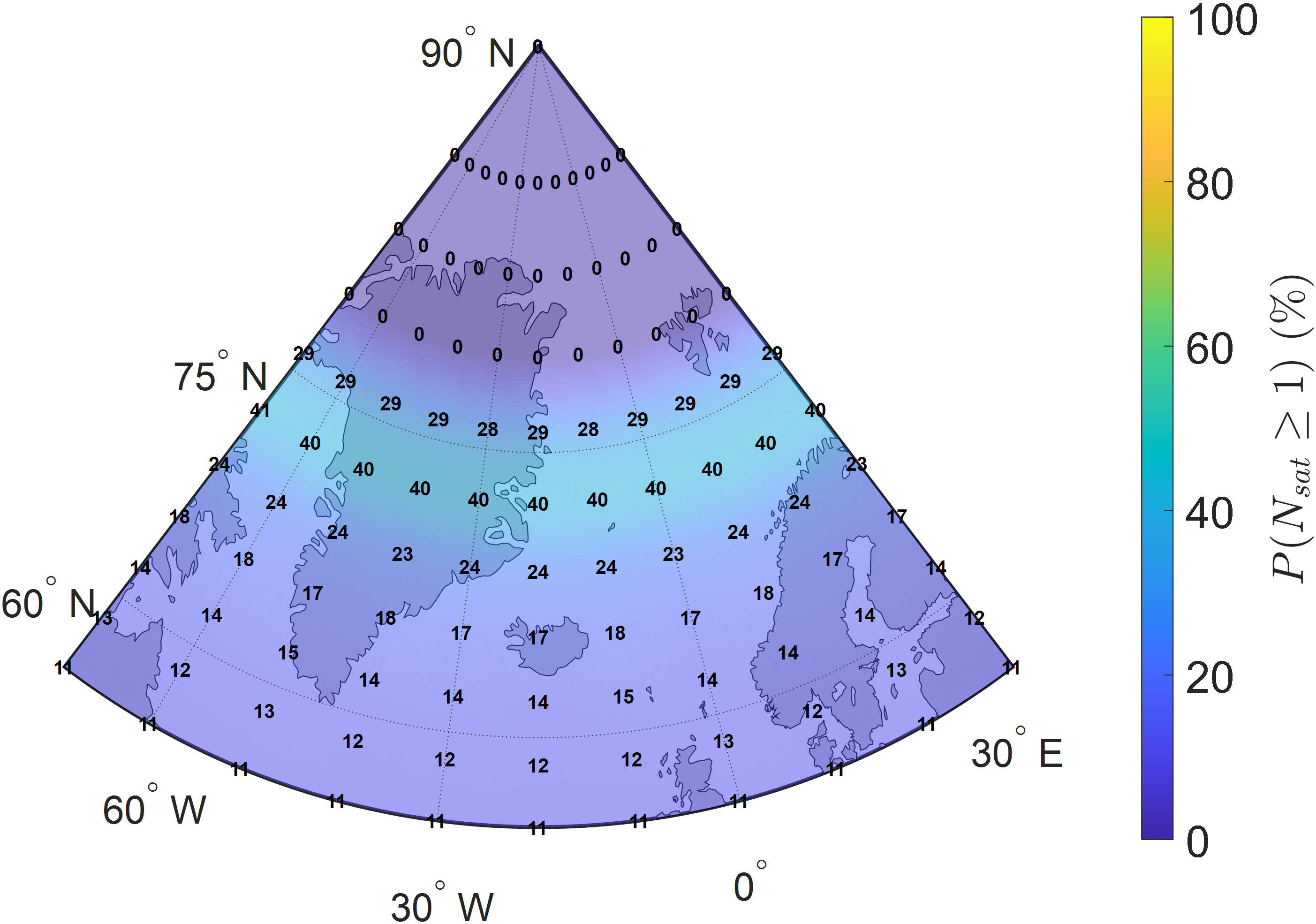}
	}
    \caption{Spatial visibility given $P(N_{sat} \geq 1)$ across the North Atlantic, considering several minimum elevation angles $\epsilon$. Results for Walker Delta ($75\degree:64/8/3$) constellation with $h = 1000$ km and (a) $\epsilon = 20\degree$, (b) $\epsilon = 40\degree$, and (c) $\epsilon = 60\degree$.} \label{fig:map_elevacion}
\end{figure*}

\begin{figure*}[!t]
\centering
\subfigure[]
{\includegraphics[width=0.45\linewidth]{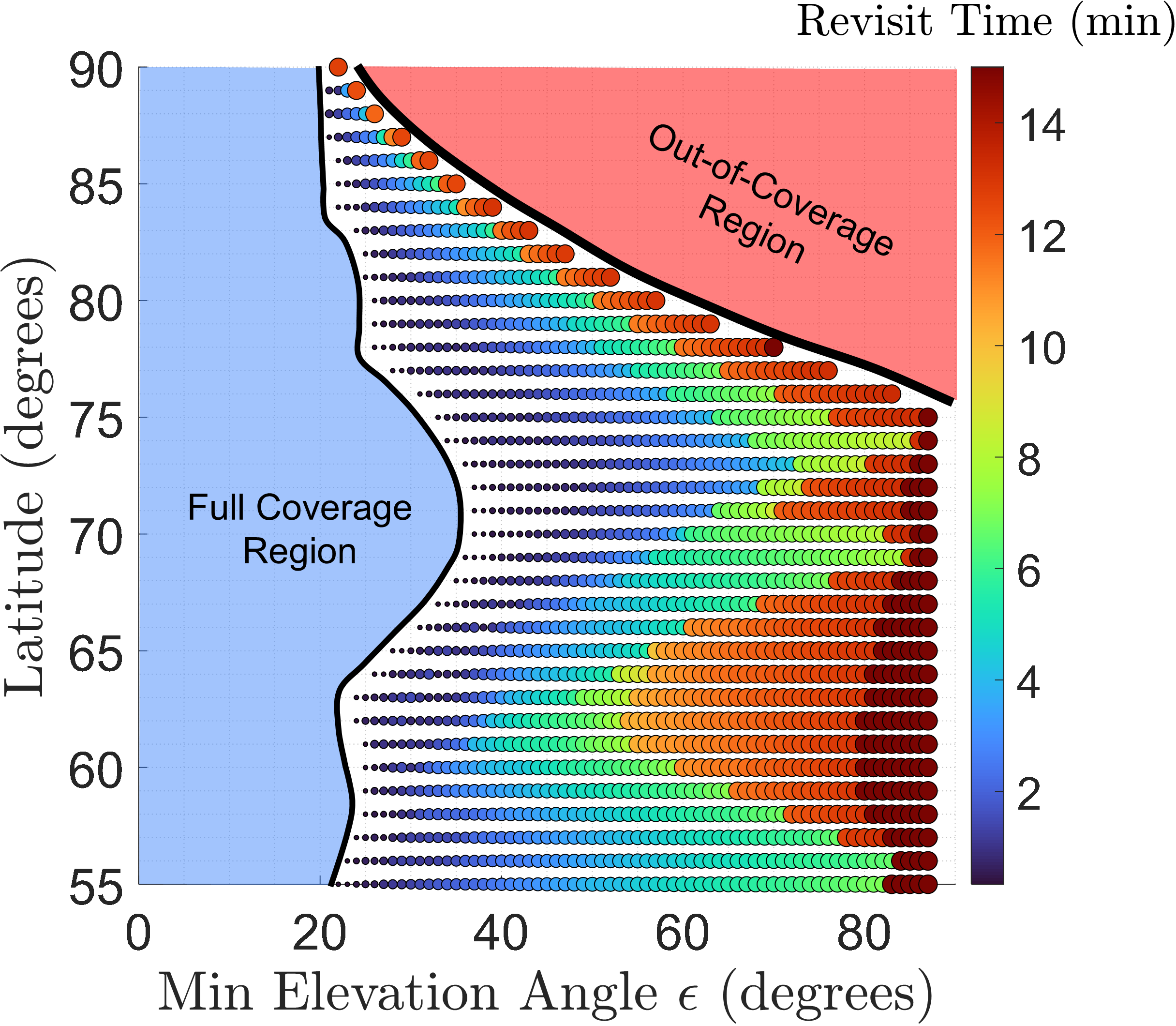}
}
\subfigure[]{\includegraphics[width=0.45\linewidth]{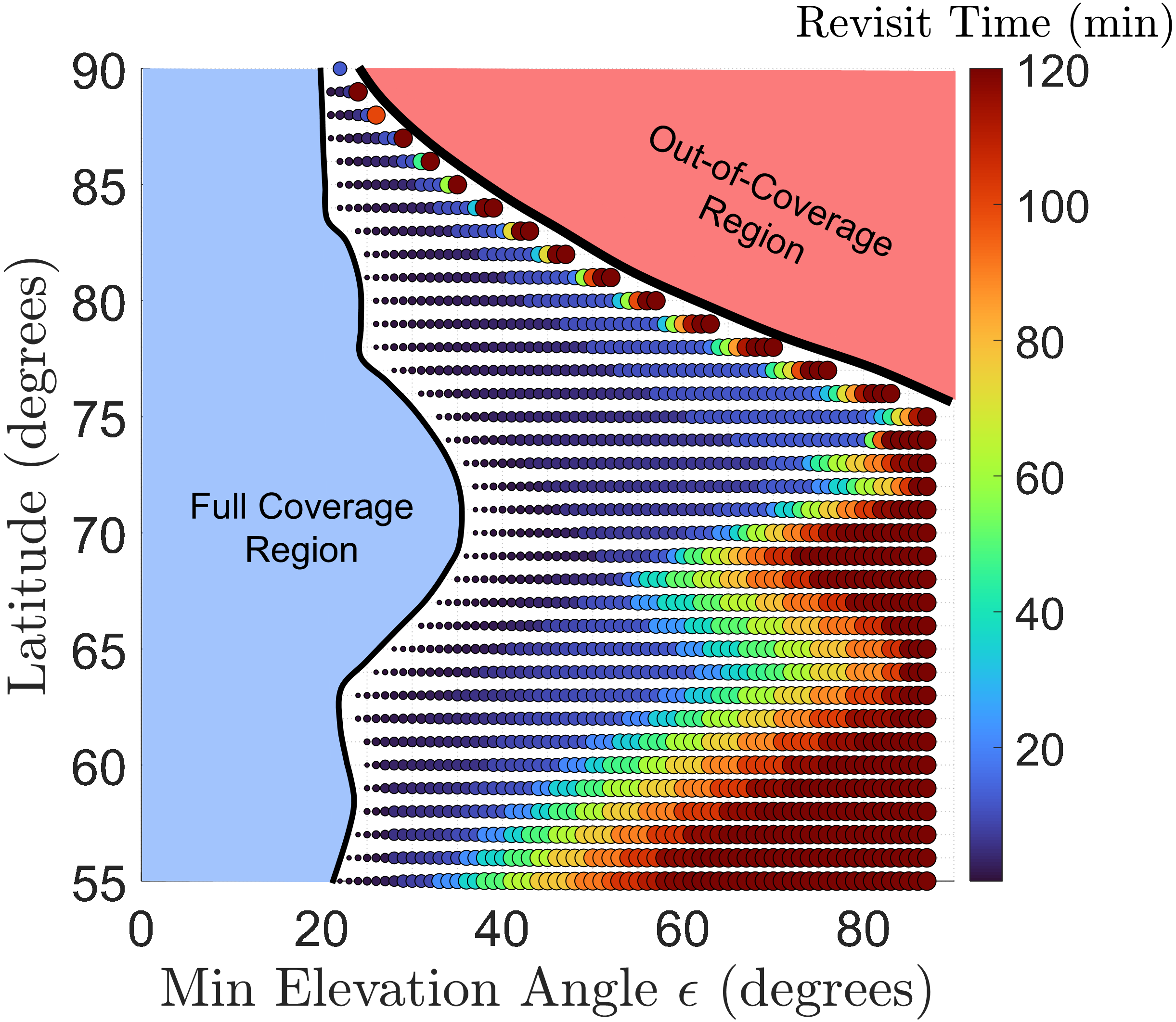}
}
\caption{(a) Median and (b) maximum revisit time $\tau$ for several latitudes and
elevation angles $\epsilon$ given a Walker Delta ($75\degree:64/8/3$) constellation with $h = 1000$ km. The full coverage region is the area where visibility to all satellites in the constellation is never lost simultaneously, while the out-of-coverage region is the area where no satellites in the constellation are ever visible.}
\label{fig:revisit_time_elevation}
\end{figure*}

\subsection{UE-Satellite Elevation Angle}

The first analysis involves the behavior of the minimum elevation angle parameter $\epsilon$. As demonstrated in the previous sections, $\epsilon$ primarily influences the coverage area of each satellite’s footprint. The larger the elevation angle, the smaller the footprint area projected onto the ground, and vice versa. To determine, from a practical standpoint, how this variable affects the constellation’s visibility, a Walker Delta constellation ($75\degree:64/8/3$) with $h = 1000$~km is simulated using the SGP4 model, considering several elevation angles. 

Fig.~\ref{fig:sweep_elevacion}(a) shows the probability of establishing visibility with one or more satellites in the constellation for latitudes between 55$\degree$ N and 90$\degree$ N, given elevation ranging from $\epsilon = 0\degree$ to $\epsilon = 90\degree$. For $\epsilon$ between $0\degree$ and $20\degree$, full coverage is observed for any latitude above $55\degree$ N, indicating that no visibility gaps are found for the footprints of the satellites under consideration. Starting at $\epsilon = 30\degree$, a degradation of $P(N_{sat} \geq 1)$ begins to be observed. At polar latitudes, visibility suffers a sudden degradation because the reduction of the footprint size, combined with a constellation inclination of $i = 75\degree$, means that the satellites never orbit over polar regions. This effect increases as the required minimum elevation angle increases, thereby reducing the maximum latitude that the constellation is capable of covering, from $90\degree$ N for $\epsilon = 20\degree$ to $78\degree$ N for $\epsilon = 70\degree$. Note that this loss of visibility at high latitudes for high minimum elevation angles is independent of the number of satellites in the constellation, since the satellite orbits are determined by the constellation’s inclination. Therefore, even a constellation with a larger number of satellites in orbit at the same inclination will still suffer the same effect. For latitudes close to the inclination angle, i.e., between $70\degree$~N and $75\degree$~N, the probability of satellite visibility reaches a maximum. For instance, for $\epsilon = 40\degree$, $P(N_{sat} \geq 1)$ reaches 98.2\% at a latitude of $72\degree$ N. This maximum value decreases significantly as the required elevation angle increases, dropping to 69.0\% at a latitude of $72\degree$~N for $\epsilon = 50\degree$, and to only 21.2\% at a latitude of $73\degree$~N for $\epsilon = 70\degree$. Improved coverage in latitudes around the constellation's inclination angle is the main reason why some satellite operators, such as SpaceX, have deployed their first shells primarily in orbits with an inclination of $53\degree$, maximizing the number of potential customers due to the geographic coverage of North America and Europe. Finally, focusing on the left side of Fig.~\ref{fig:sweep_elevacion}(a), for latitudes below $70\degree$~N, a gradual decrease in visibility is observed for all possible elevation angles $\epsilon$. As latitude decreases from $70\degree$~N toward the equator, the surface area to be covered by the Walker Delta constellation increases. This is due to Earth's larger circumference at lower latitudes, which reduces satellite density per unit area and consequently degrades visibility.

Fig.~\ref{fig:sweep_elevacion}(b) shows the isolines for $P(N_{sat} \geq 1)$ at 99\%, 90\%, 75\%, 50\%, 25\%, 10\%, and 1\% in a two-dimensional space defined by latitude and the minimum elevation angle $\epsilon$. As shown in Fig.~\ref{fig:sweep_elevacion}(a), full coverage is observed for $\epsilon \leq 20\degree$ at latitudes above $55\degree$ N. This visibility gradually deteriorates as $\epsilon > 20\degree$, although it is slightly better in the geographical range between $70\degree$ N and $75\degree$ N compared to other latitudes. Note the triangle formed in the upper right corner, which represents the region where coverage is null as the footprint does not reach polar regions. Figs.~\ref{fig:map_elevacion}(a)–(c) show the spatial visibility over the North Atlantic for three specific satellite constellation examples, with (a) $\epsilon = 20\degree$, (b) $\epsilon = 40\degree$, and (c) $\epsilon = 60\degree$, respectively. Focusing on longitudes between $70\degree$ W and $30\degree$ E, which primarily cover Greenland, the North Atlantic, and Scandinavia, it can be seen that the proposed constellation can provide seamless coverage assuming $\epsilon = 20\degree$ (see Fig.~\ref{fig:map_elevacion}(a)). However, when $\epsilon$ is increased to $40\degree$~(see Fig.~\ref{fig:map_elevacion}(b)), coverage remains above 90\% only at latitudes between approximately $70\degree$~N and $75\degree$~N, which corresponds only to certain parts of Greenland and the North Atlantic, and northern Scandinavia. In Southern Scandinavia, the probability of establishing visibility drops to below 50\%, and in the Arctic regions this probability plummets to zero. In a more extreme case, where $\epsilon = 60\degree$ (see Fig.~\ref{fig:map_elevacion}(c)), visibility reaches only 40\% in a narrow strip along the North Atlantic, while it drops to 17–24\% in northern Scandinavia and to 11–13\% in southern Scandinavia.

Not only is the probability of visibility critical to the design of the constellation, but so is the revisit time of the satellites that form the constellation. If $P(N_{sat} \geq 1) \neq 100\%$, there will be temporary events $n$ during which a given ground coordinates cannot establish connectivity with any satellite in the constellation. Assuming use cases in which certain disconnections of duration $\tau_n$, i.e., the revisit time, are tolerable, the ideal approach is to design the constellations such that the distribution of $\tau$ is as uniform as possible, thereby minimizing the maximum time without connectivity to the constellation. Based on the parametric analysis as a function of latitude and minimum elevation angle, Fig.~\ref{fig:revisit_time_elevation}(a) shows the median value of the $\tau$ distribution across the two-dimensional latitude-elevation domain. The region on the left represents the domain where $P(N_{sat} \geq 1) = 100\%$, and therefore, the effective revisit time is zero. Equivalently, the out-of-coverage region represents the domain where $P(N_{sat} \geq 1) = 0\%$, implying an effective revisit time tending to infinity. Looking at the rest of the domain, it can be seen that for most of the cases the median value of $\tau$ is bounded below 15 minutes, except for constellations with $\epsilon > 80\degree$, where the color map has been capped at 15 minutes since the values rise in this region due to the reduced area of the footprint projection. Similar to the probability of visibility, there is a region between $70\degree$ N and $75\degree$ N latitude where the median remains below 5 minutes even for elevation angles $\epsilon$ of up to $60\degree$ due to the constellation’s $75\degree$ inclination. Additionally, a region between $55\degree$ N and $60\degree$ N latitude is observed with a median value similar to that found in the region between $70\degree$ N and $75\degree$ N. This might lead one to think that both regions are covered equally well. However, if we consider the maximum value of $\tau$ (see Fig.~\ref{fig:revisit_time_elevation}(b)), that is, the maximum time without connectivity given the dynamics of the satellites and the constellation’s orbits, the maximum revisit time increases progressively as we move down in latitude. This fact demonstrates how a medium-sized constellation ($T = 64$) struggles to maintain connectivity as we move towards the Earth’s equator. Looking at the distribution of the maximum $\tau$, in most cases the maximum revisit time is bounded below 30 minutes, being this particularly true for higher latitudes across almost any range of elevation angles, or for lower latitudes provided that $\epsilon$ is bounded. Otherwise, the maximum revisit time quickly exceeds 120 minutes (see the red samples in Fig.~\ref{fig:revisit_time_elevation}(b)).

This subsection clearly illustrates the impact of considering different minimum elevation angle requirements $\epsilon$. Thus, this analysis highlights the need to establish, at an early stage of satellite constellation design, the required elevation angle between ground stations and satellites, in order to define minimum connectivity requirements and perform optimal network sizing. As an illustrative example, and based on the previous analysis, the same satellite constellation may tentatively have 100\% visibility between ground stations and satellites for latitudes above 55$\degree$ N, assuming $\epsilon = 20\degree$, whereas if we assume $\epsilon = 60\degree$, visibility in certain regions may decrease to $\sim 10\%$ or even be zero at the poles due to a narrower footprint. This selection will be determined by multiple factors, such as ground station hardware and satellites, antenna directivity, expected path loss, and minimum probability of Line-of-Sight under the assumption of potential obstacles on the ground, as described in Section~III.

\subsection{Inclination}

\begin{figure*}[!b]
	\centering
	\subfigure[]{\includegraphics[width=0.59\linewidth]{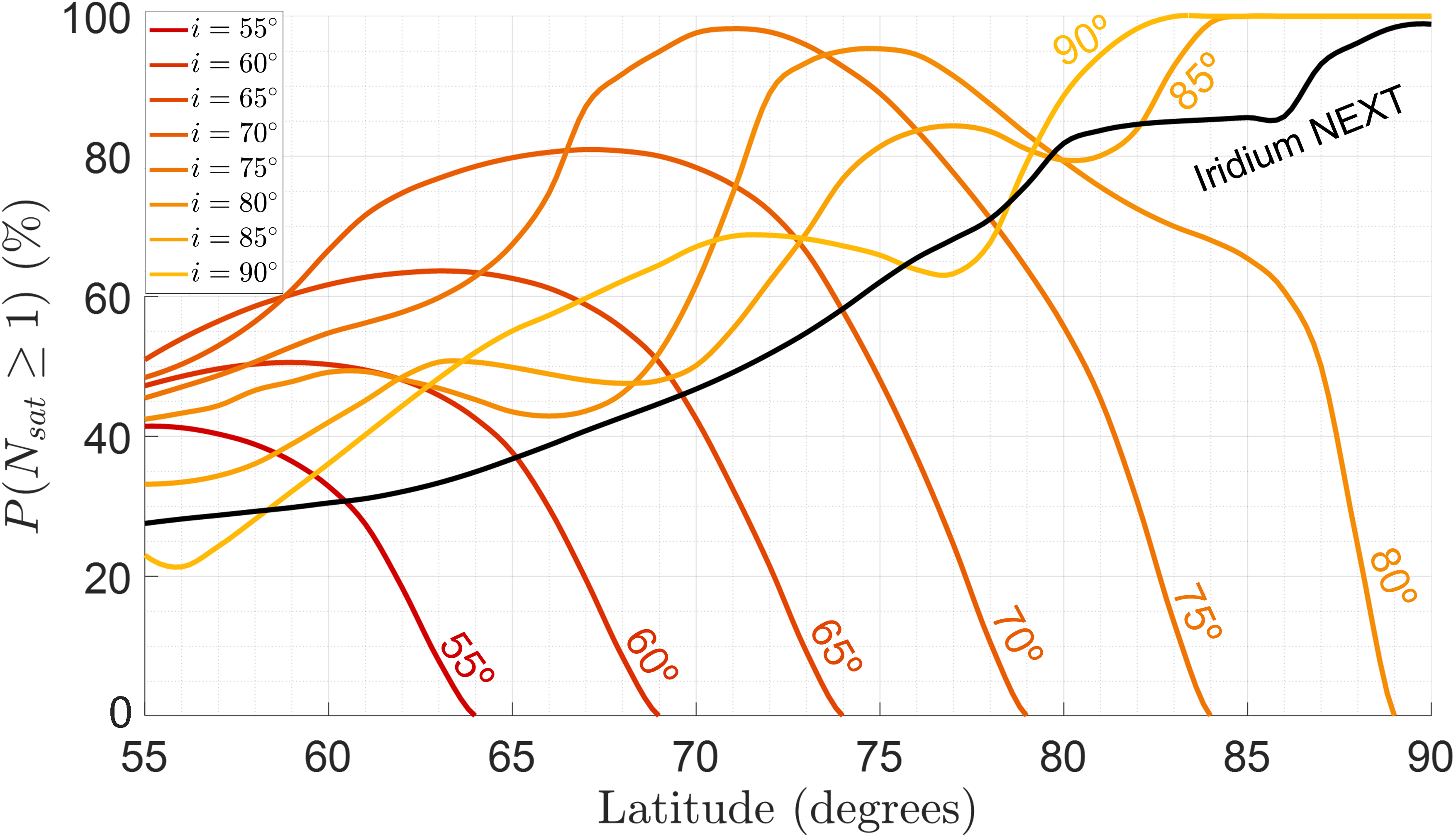}
	} 
    \subfigure[]{\includegraphics[width=0.39\linewidth]{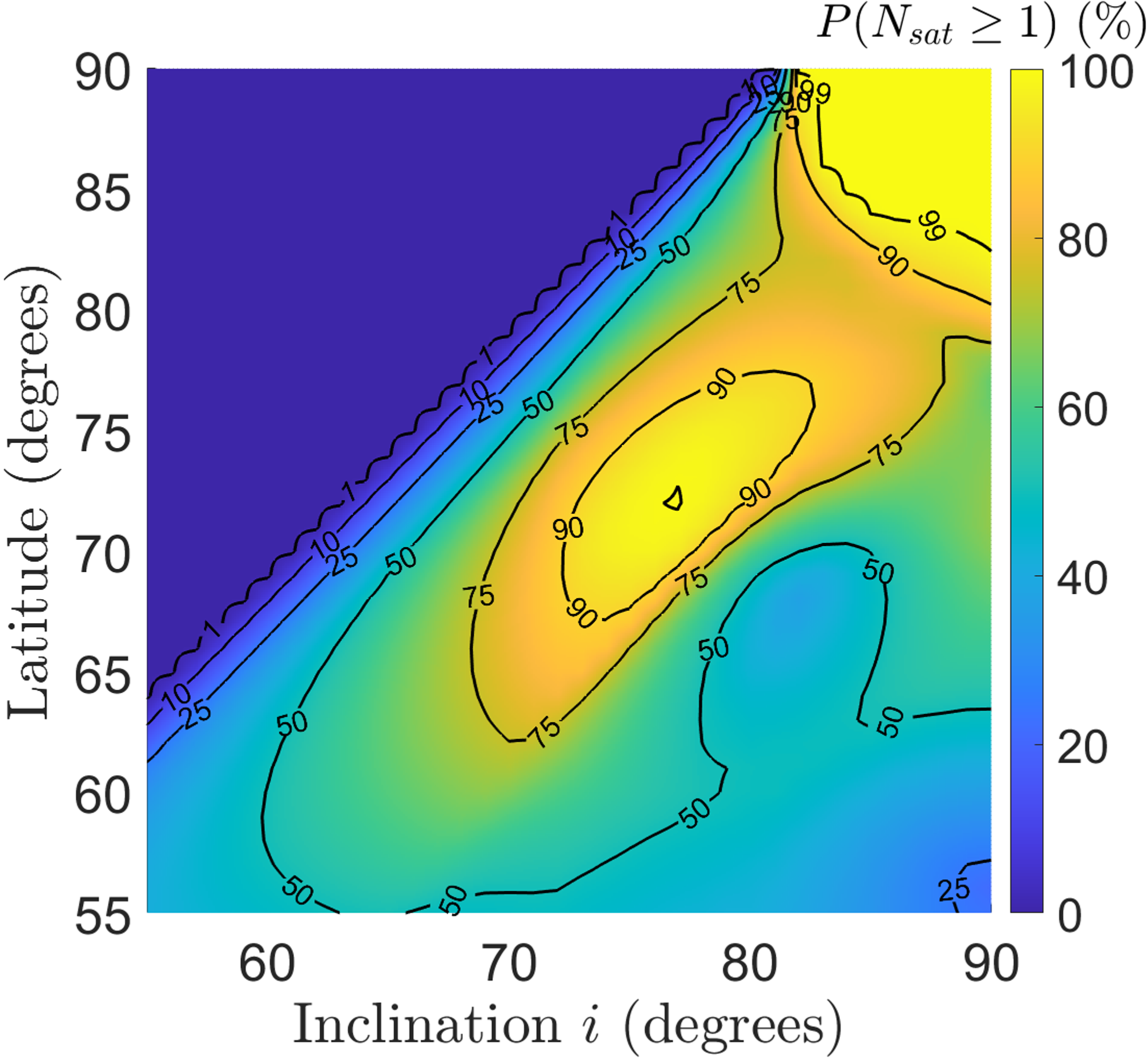}
	}
    \caption{(a) Probability of establishing visibility with at least one satellite in the LEO constellation for several latitudes and inclination angles $i$, and (b) isolines for $P(N_{sat} \geq 1)$ at 99\%, 90\%, 75\%, 50\%, 25\%, 10\%, and 1\%. Results for Walker Delta ($i:64/8/3$) constellation with $h = 1000$ km and $\epsilon = 40\degree$. Iridium NEXT constellation is also shown for comparison purposes.} \label{fig:sweep_inclination_probability}
\end{figure*}

\begin{figure}
\centering
\includegraphics[width=1\linewidth]{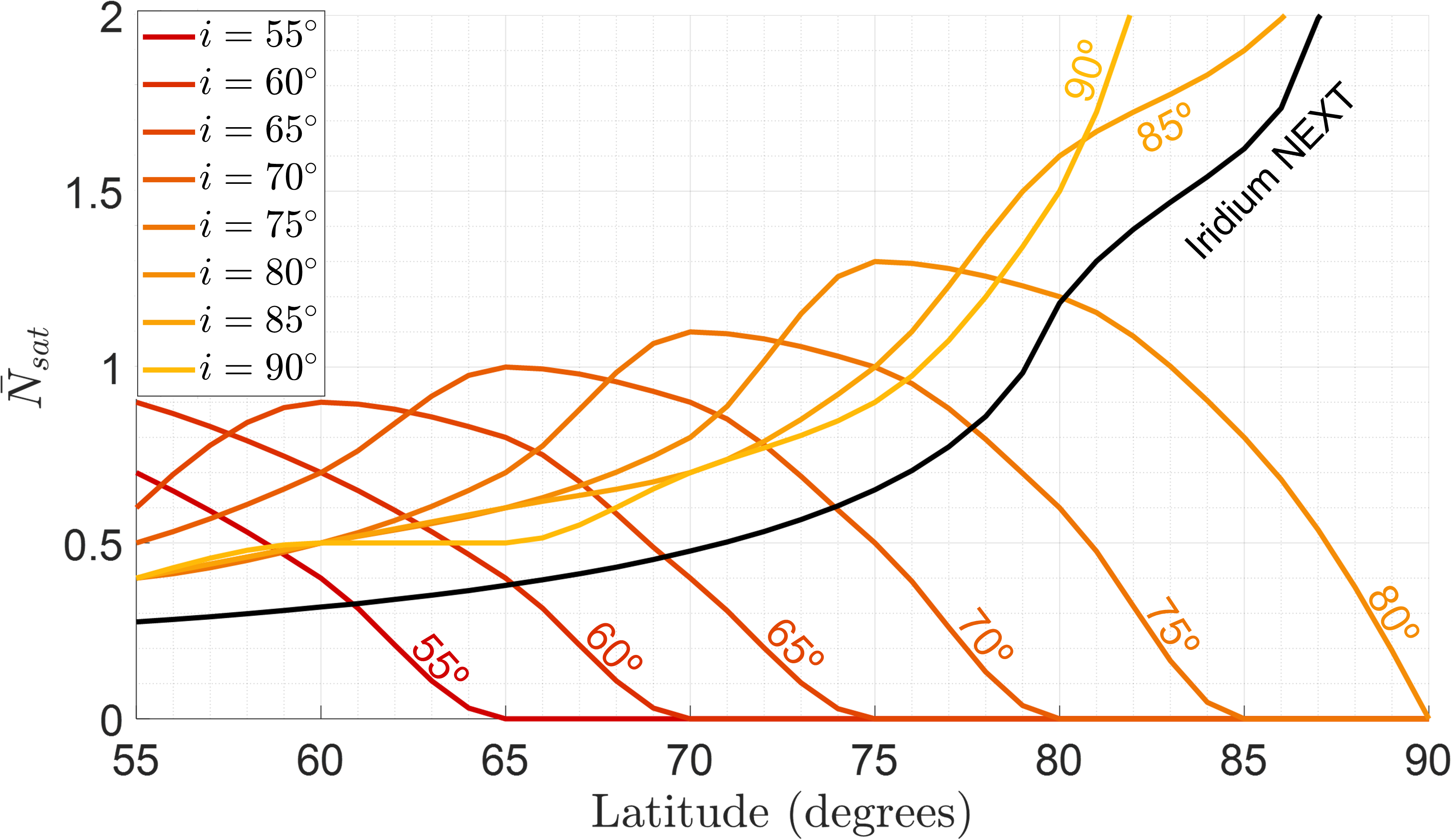}
\caption{Average number of satellites within the ground station visibility range in the LEO constellation for several latitudes and inclination angles $i$. Results for Walker Delta ($i:64/8/3$) constellation with $h = 1000$ km and $\epsilon = 40\degree$. Iridium NEXT constellation is also shown for comparison purposes.}
\label{fig:sweep_inclination_average}
\end{figure}

\begin{figure*}[t]
	\centering
	\subfigure[{\footnotesize $i = 55\degree$}]{\includegraphics[width=0.32\linewidth]{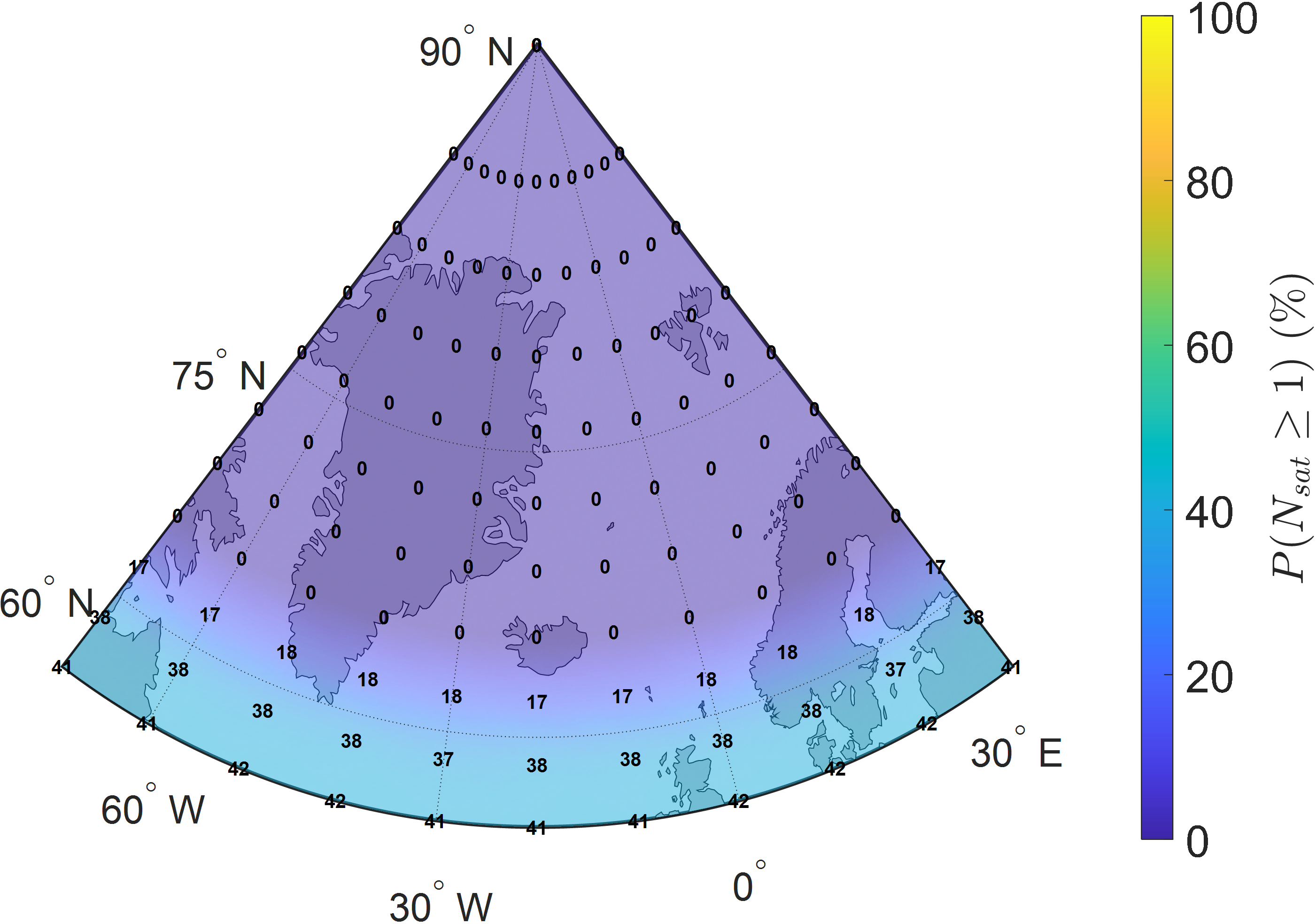}
	} 
    \subfigure[{\footnotesize $i = 75\degree$}]{\includegraphics[width=0.32\linewidth]{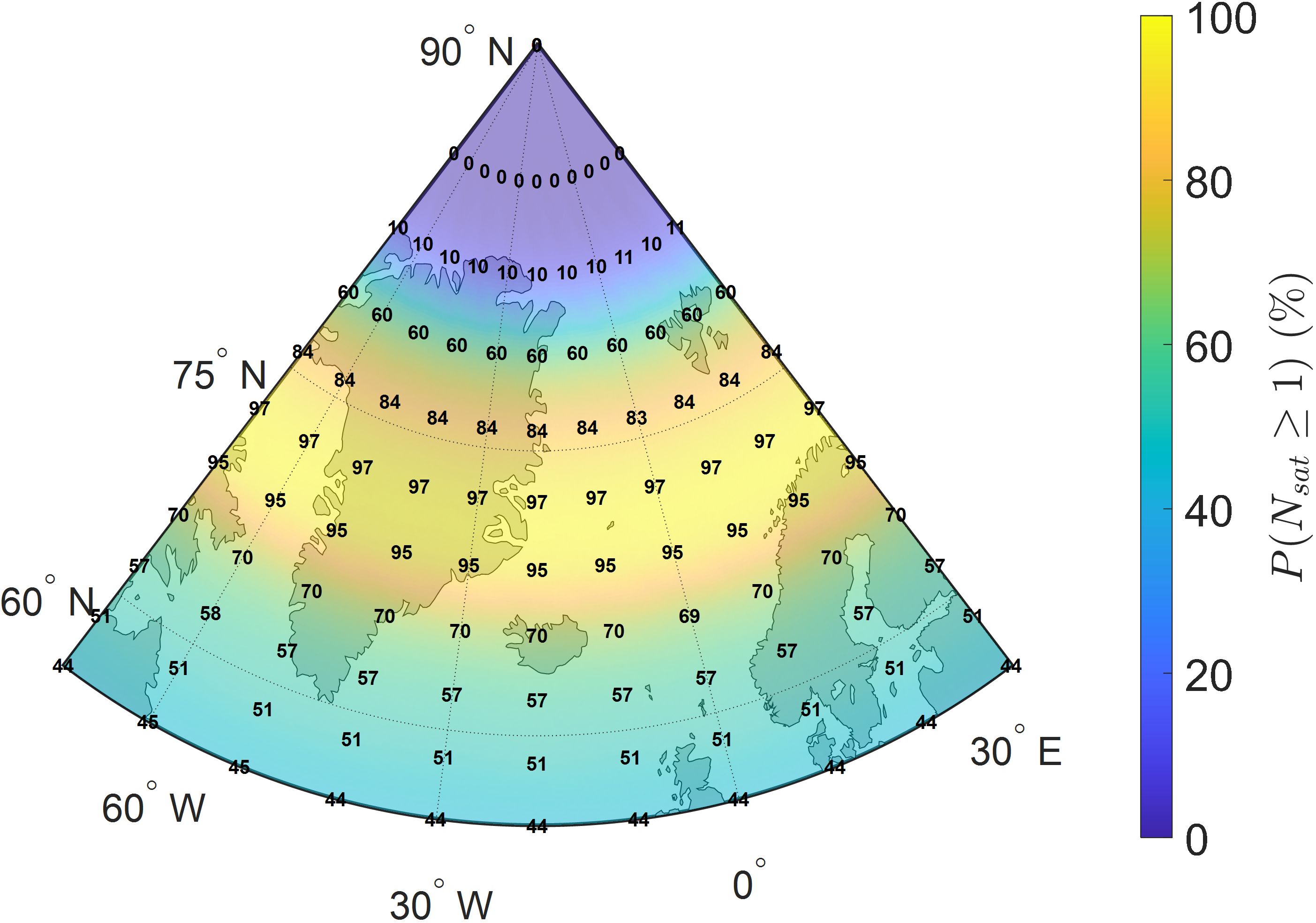}
	}
    \subfigure[{\footnotesize $i = 90\degree$}]{\includegraphics[width=0.32\linewidth]{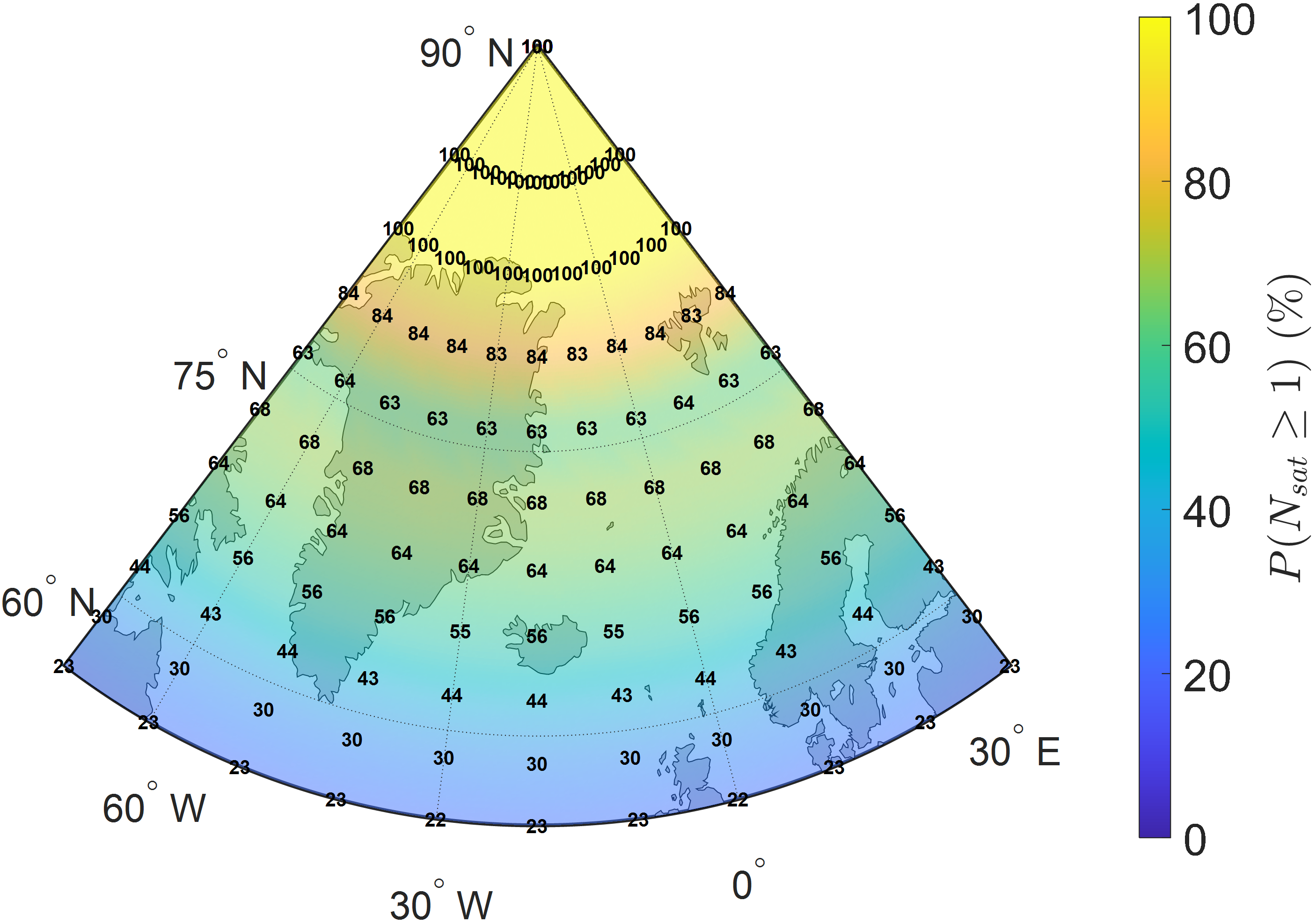}
	}
    \caption{Spatial visibility given $P(N_{sat} \geq 1)$ across the North Atlantic, considering several inclination angles $i$. Results for Walker Delta ($i:64/8/3$) constellation with $h = 1000$ km and $\epsilon = 40\degree$ and (a) $i = 55\degree$, (b) $i = 75\degree$, and (c) $i = 90\degree$.} \label{fig:map_inclination}
\end{figure*}

\begin{figure*}[!t]
\centering
\subfigure[]
{\includegraphics[width=0.45\linewidth]{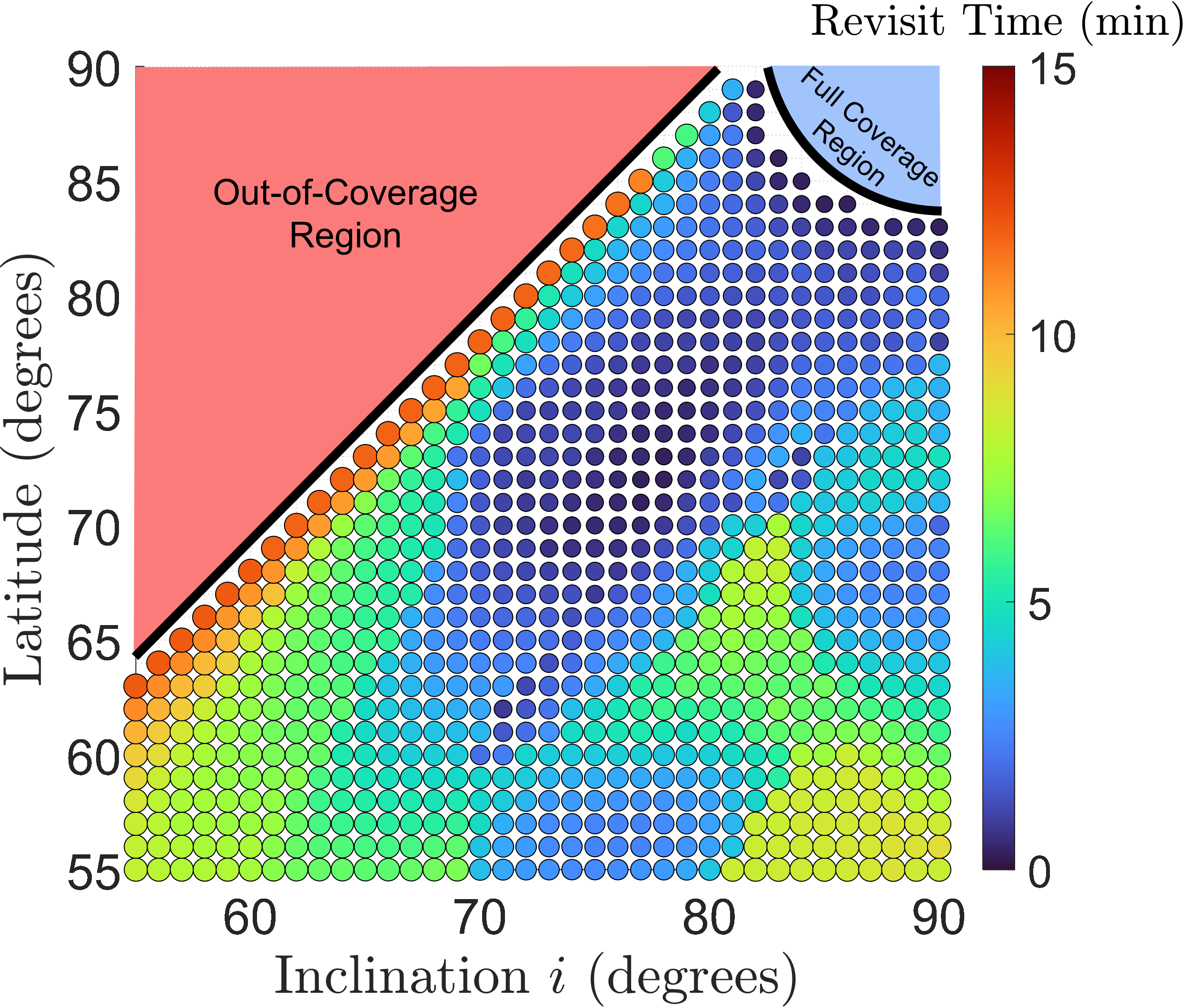}
}
\hspace{0.5cm}
\subfigure[]{\includegraphics[width=0.45\linewidth]{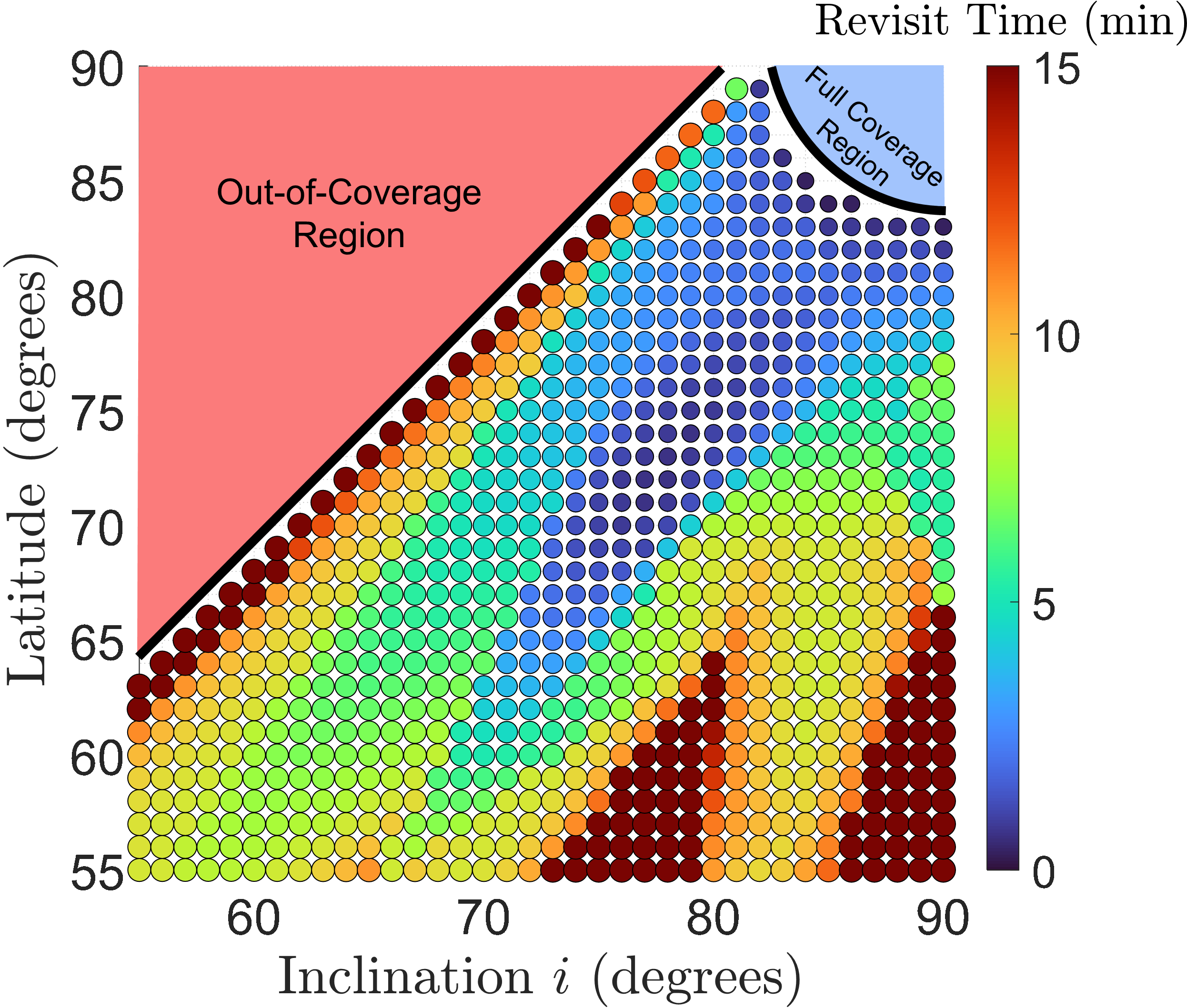}
}
\caption{(a) Median and (b) maximum revisit time $\tau$ for several latitudes and
inclination angles $i$ given a Walker Delta ($i:64/8/3$) constellation with $h = 1000$ km and $\epsilon = 40\degree$. The full coverage region is the area where visibility to all satellites in the constellation is never lost simultaneously, while the out-of-coverage region is the area where no satellites in the constellation are ever visible.}
\label{fig:revisit_time_inclination}
\end{figure*}

The second analysis involves the assessment of the inclination $i$ and how it models the satellite link visibility on Earth. As explained in Section~I and at the beginning of this Section~IV, commercial deployments adopt a primarily binary approach: (i) either low-inclination constellations ($i < 55\degree$), or (ii) polar orbit constellations ($i \approx 90\degree$). This subsection addresses intermediate cases to determine which orbital configurations may be of interest for improving coverage in the North Atlantic with a reduced number of satellites. To this end, and based on the assumptions and analyses of previous subsections, a Walker Delta constellation ($i:64/8/3$) with $h = 1000$~km and $\epsilon = 40\degree$ is considered, where the orbital inclination is variable. This minimum elevation angle is chosen as a trade-off between the visibility discussed in the subsection \textit{UE-Satellite Elevation Angle} and factors such as the minimum elevation angle allowed by user terminals, the probability of line-of-sight, or losses due to atmospheric attenuation for low elevation angles described throughout Section~III. The constellation is simulated using the SGP4 model, considering multiple inclinations as well as multiple latitudes on Earth.

Fig.~\ref{fig:sweep_inclination_probability}(a) shows the probability of visibility with at least one satellite for several orbital inclinations ranging from $55\degree$ to $90\degree$ across multiple geographic latitudes. Examining the probability distributions reveals a consistent pattern across them. The lower the orbital inclination, the sooner visibility is lost at high latitudes, with $P(N_{sat} \geq 1)$ becoming null at latitudes $9\degree$ above the orbital inclination. This exact value is determined by the elevation angle $\epsilon = 40\degree$, increasing for low $\epsilon$ values and decreasing for high $\epsilon$ values. In any case, the pattern is similar regardless of $\epsilon$. Based on the previous statement, any inclination greater than $i > 81\degree$ is capable of providing coverage at the poles. Regarding the maximum $P(N_{sat} \geq 1)$ reached for each configuration, this is achieved at values where the latitude range is close to the orbital inclination. Additionally, this maximum tends to decrease for lower inclinations because the area to be covered, given the Earth’s circumference, increases at lower latitudes, so a fixed number of satellites cannot meet the same demand at polar latitudes as at equatorial latitudes. For instance, for an inclination $i = 90\degree$, $P(N_{sat} \geq 1) = 100\%$ is achieved at latitudes larger than $83\degree$ N. For $i = 70\degree$, the maximum probability of visibility is reached at a latitude of $67\degree$ N, with $P(N_{sat} \geq 1) = 80.9\%$. Finally, for $i = 60\degree$, the maximum is reached at a latitude of $59\degree$ N, with $P(N_{sat} \geq 1) = 50.5\%$. Note how the maximum decreases along with the inclination, as explained earlier. From a two-dimensional perspective, Fig.~\ref{fig:sweep_inclination_probability}(b) shows the isolines for $P(N_{sat} \geq 1)$ at 99\%, 90\%, 75\%, 50\%, 25\%, 10\%, and 1\% in a two-dimensional space defined by latitude and the inclination $i$. Following the discussion in the previous Fig.~\ref{fig:sweep_inclination_probability}(a), the diagonal formed between latitude and inclination defines the optimal visibility region, indicating the correlation between the two variables. The upper left region clearly shows the area without coverage due to satellites that do not cover high latitudes, given a specific inclination. This region is bounded by a line with a $45\degree$ slope, confirming that the maximum covered latitude remains constant at $9\degree$ latitude above the orbital inclination. Regarding the lower right region, note how $P(N_{sat} \geq 1)$ decreases relative to the values observed along the diagonal, indicating the difficulty of establishing visibility due to the lower density of satellites at low latitudes. To illustrate this effect, Fig.~\ref{fig:sweep_inclination_average} shows the average number of satellites within visibility range ($\bar{N}_{sat}$) from the ground for the analysis in Fig.~\ref{fig:sweep_inclination_probability}(a). Comparing the two figures reveals a strong correlation between $P(N_{sat} \geq 1)$ and $\bar{N}_{sat}$. For instance, for polar orbits ($i \approx 90\degree$), the average number of satellites at polar latitudes consistently exceeds $\bar{N}_{sat} > 2$, whereas for orbits with $i \approx 60\degree$, $\bar{N}_{sat}$ does not even reach an average of $\bar{N}_{sat} = 1$ in the best case scenario. As discussed in previous sections, these visibility limitations are inherent to a modest constellation in terms of the number of satellites, in contrast to mega-constellations, where, regardless of the orbital configuration, seamless visibility can be achieved at latitudes below the inclination angle. However, the main goal of the previous results is to understand satellite dynamics, which are more noticeable in small to medium constellations such as the one analyzed. For reference, Iridium NEXT metrics regarding visibility probability and average number of satellites are included in Figs.~\ref{fig:sweep_inclination_probability}(a) and~\ref{fig:sweep_inclination_average}, respectively. These curves resemble those of the Walker Delta constellation with inclinations close to $i = 90\degree$ due to Iridium NEXT’s near-polar inclination ($i = 86.4\degree$) and a similar number of satellites (64 for Walker Delta vs. 66 for Iridium NEXT). In both figures, visibility is slightly lower for Iridium NEXT due to the constellation’s lower altitude (1,000 km for Walker Delta vs. 780 km for Iridium NEXT), which results in a slight reduction in the footprint on Earth. A detailed analysis of this operational Iridium NEXT constellation is presented at the end of this Section~IV.

Figs.~\ref{fig:map_inclination}(a)–(c) show three examples of the coverage achieved by different constellations over the North Atlantic, illustrating the spatial visibility for inclinations of (a) $i = 55\degree$, (b) $i = 75\degree$, and (c) $i = 90\degree$, respectively. Fig.~\ref{fig:map_inclination}(a) illustrates a low-inclination deployment, which clearly compromises coverage in the North Atlantic and Arctic regions, with $P(N_{sat} \geq 1)$ values reaching only 42\% in regions far south of Scandinavia. Looking at Fig.~\ref{fig:map_inclination}(b), we can see that increasing the inclination to $i = 75\degree$ generally improves visibility in this region, reaching peaks of 97\% visibility that decrease to 44\% in southern Scandinavia. Note that the values at $55\degree$ N latitudes are similar for both $i = 55\degree$ and $i = 75\degree$, implying that increasing the inclination offers only advantages as long as the region of interest comprises exclusively latitudes above $55\degree$ N. Even so, and according to the proposed configuration, the inclination remains too low to provide visibility in the Arctic. To solve this problem, one must opt for configurations such as the one shown in Fig.~\ref{fig:map_inclination}(c) with $i = 90\degree$, where coverage in the Arctic is complete. As a trade-off, visibility gradually decreases as we move toward lower latitudes, dropping to just 22\% in southern Scandinavia.

In cases where there is no seamless visibility, it is of interest to analyze the revisit time in the same manner as it has been carried out in Figs.~\ref{fig:revisit_time_elevation}(a) and~\ref{fig:revisit_time_elevation}(b) for the minimum elevation angle $\epsilon$. Figs.~\ref{fig:revisit_time_inclination}(a) and~\ref{fig:revisit_time_inclination}(b) show the median and maximum revisit times, respectively, for multiple geographic latitudes and constellation inclinations $i$. Focusing on Fig.~\ref{fig:revisit_time_inclination}(a), we can see the red region where the revisit time does not apply since the visibility null, and the blue region where the revisit time is zero because there is always at least one available satellite. In the rest of the domain, it can be seen that along the diagonal, i.e., the region where latitude and inclination are similar, the lowest revisit times are achieved, while these tend to increase as we move away from that diagonal. In the upper right quadrant of the diagonal, most median revisit times are clustered below 5 minutes, while in the lower right quadrant these values significantly exceed 5 minutes. This is due to the satellite density decreasing considerably as ground station move to lower latitudes. With regard to the maximum revisit time (see Fig.~\ref{fig:revisit_time_inclination}(b)), it behaves similarly to the median, with the optimal values being those near the diagonal where the latitude is similar to the inclination $i$. Note that in some cases, the maximum revisit time increases significantly to 15 minutes or more when inclinations larger than $i = 73\degree$ are used and coverage is intended for lower latitudes below $\sim60\degree$ $-$ $65\degree$N.

This subsection highlights the importance of choosing an appropriate inclination based on the requirements of ground stations and their locations. While constellations with global coverage, such as Starlink or OneWeb, require hundreds and thousands of satellites, constellations with regional coverage requirements, such as those covering the North Atlantic, require a smaller number of satellites to provide full regional visibility at high latitudes. The analysis presented here examines different operational ranges based on inclination for high-latitude regions and the satellite dynamics for multiple inclinations. The results show the need for constellations with $i \gtrsim70\degree$ if one wishes to provide broad coverage in the North Atlantic with a $(i:64/8/3)$ Walker Delta-type constellation for $\epsilon = 40\degree$ and $h = 1000 \textrm{ km}$, requiring an even higher inclination if one wishes to improve visibility in Arctic regions for a constellation with the same number of satellites. Note that this recommendation applies to the geometric constellation scenario described throughout this section. However, as a general guideline, relaxing the constraints on the minimum elevation angle $\epsilon$, and increasing the constellation altitude $h$ and/or the number of satellites, would allow for a reduction in orbital inclination while maintaining visibility in the North Atlantic.

\begin{figure*}[!t]
\centering
\subfigure[]
{\includegraphics[width=0.4\linewidth]{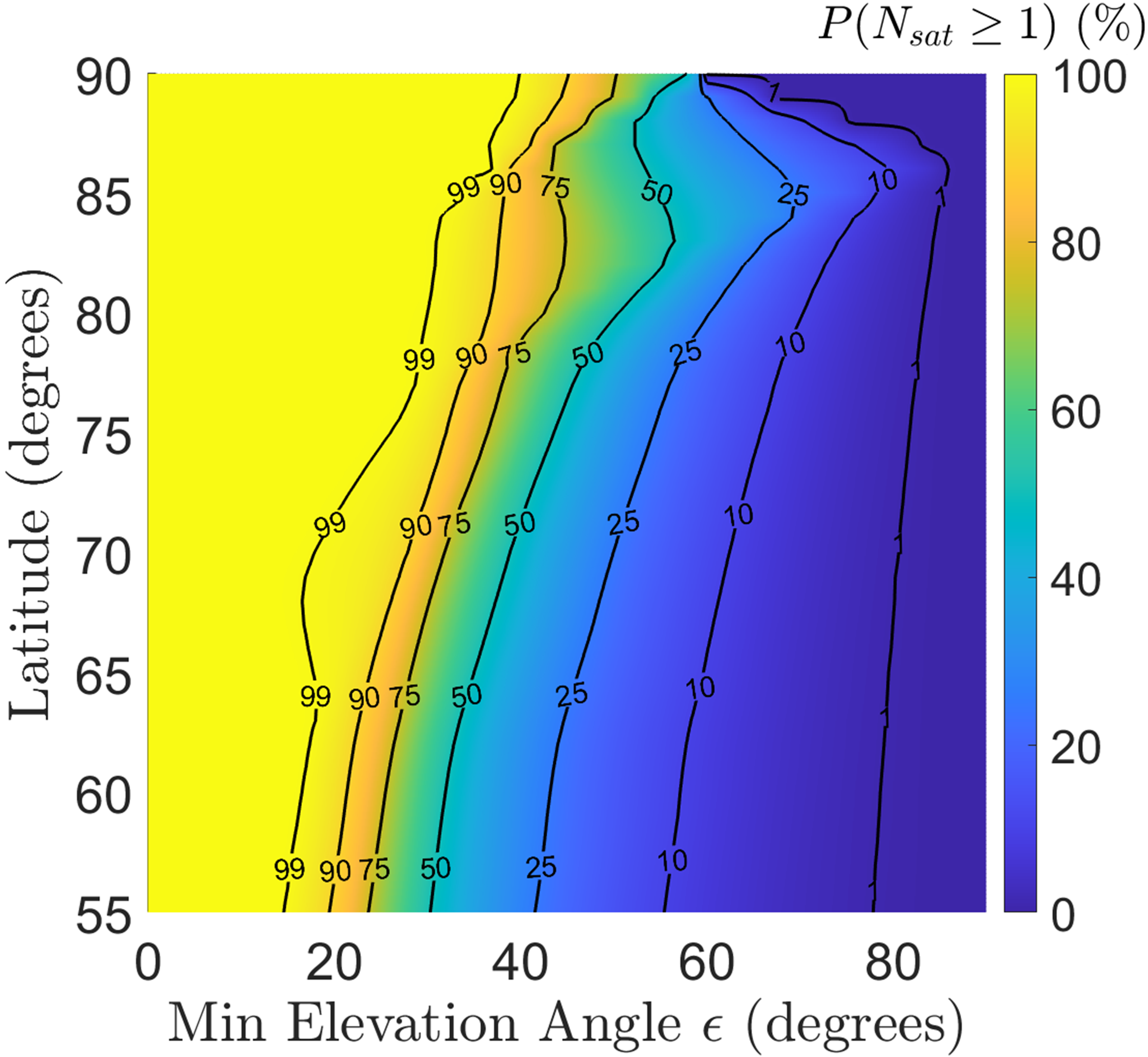}
}
\hspace{0.5cm}
\subfigure[]{\includegraphics[width=0.45\linewidth]{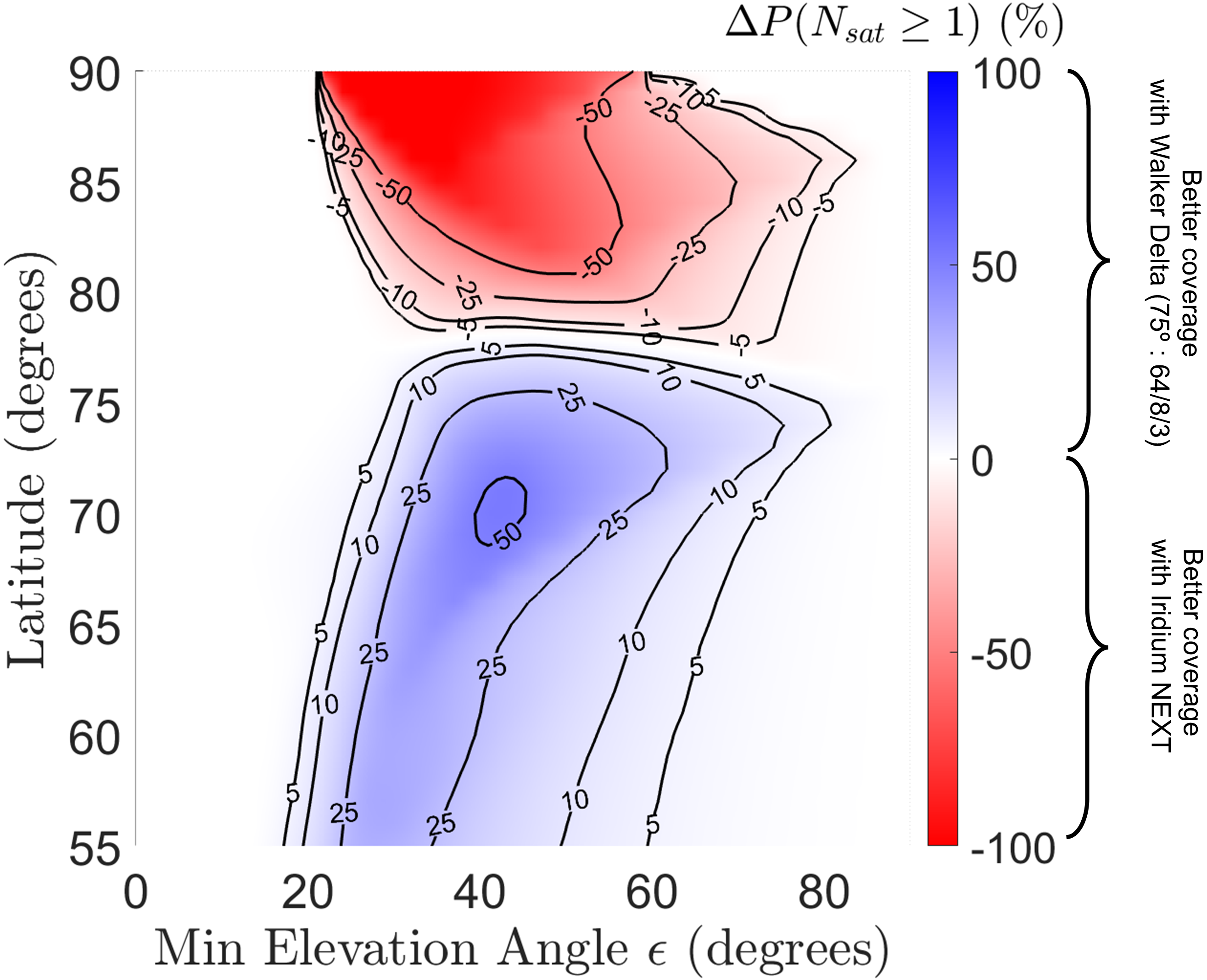}
}
\caption{(a) Probability of establishing visibility with at least one satellite in the Iridium NEXT Constellation for several latitudes and elevation angles $\epsilon$. (b) Comparison of visibility probability between the Iridium NEXT constellation and the Walker Delta ($75\degree:64/8/3$) constellation with $h = 1000$ km. Blue tones indicate better visibility for the Walker Delta constellation, while red tones indicate better visibility for the Iridium NEXT constellation.}
\label{fig:iridium_probability}
\end{figure*}

\begin{figure*}[!t]
\centering
\subfigure[]
{\includegraphics[width=0.43\linewidth]{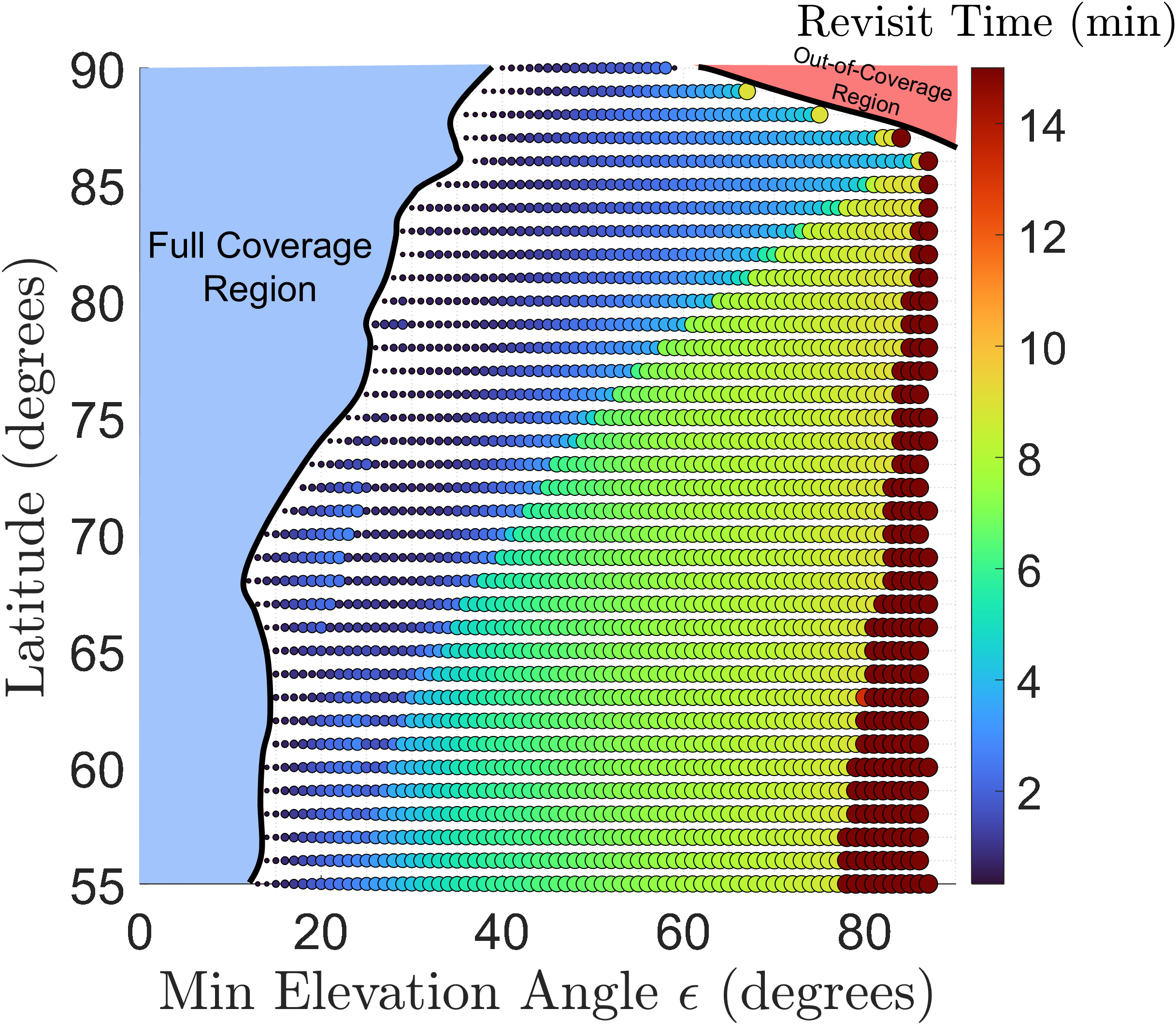}
}
\hspace{0.4cm}
\subfigure[]{\includegraphics[width=0.43\linewidth]{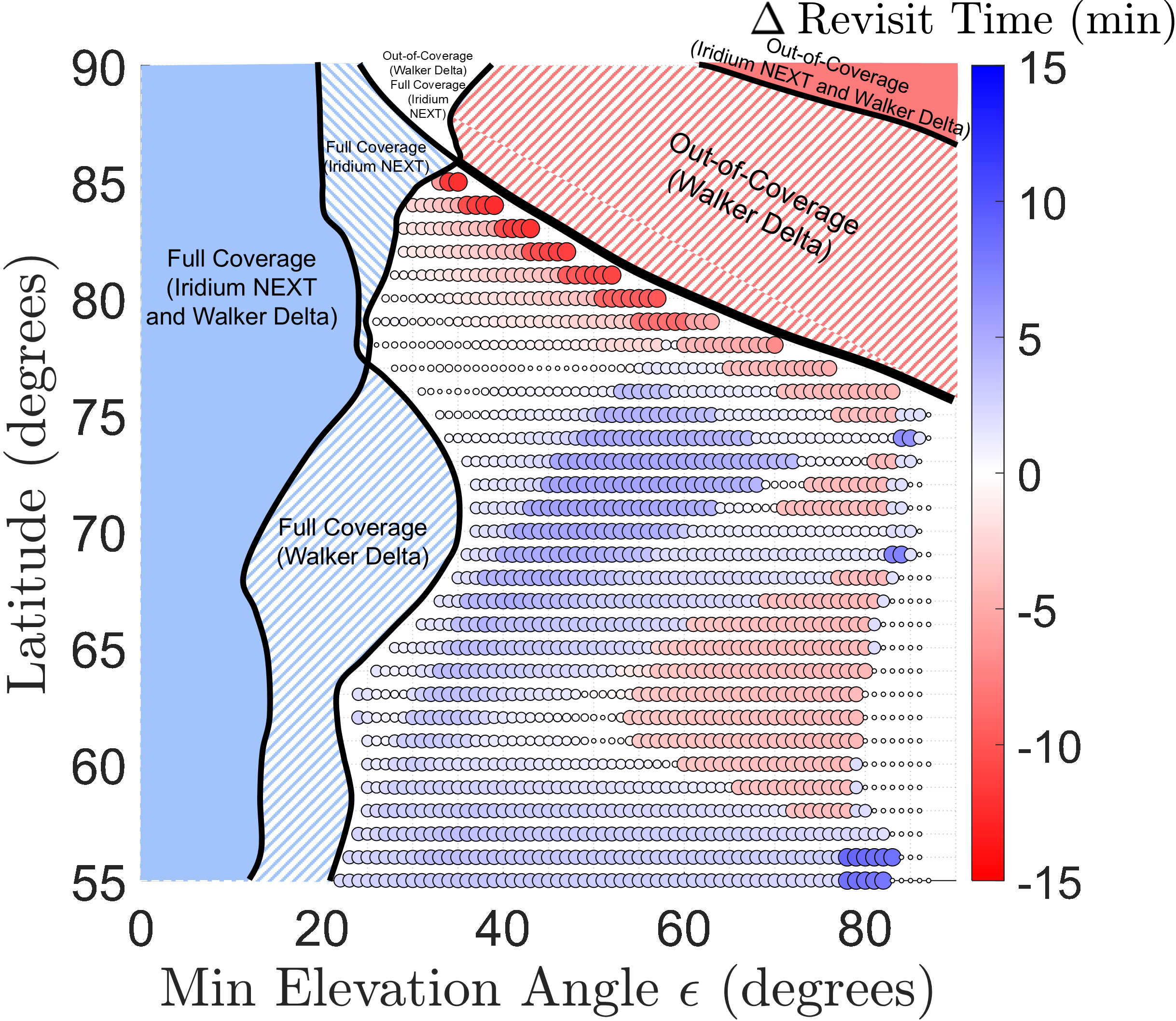}
}
\caption{(a) Median revisit time $\tau$ for several latitudes and
elevation angles $\epsilon$ given the Iridium NEXT constellation. (b) Comparison of median revisit time between the Iridium NEXT constellation and the Walker Delta ($75\degree:64/8/3$) constellation with $h = 1000$ km. Blue dots indicate a lower median revisit time for the Walker Delta constellation, while red tones indicate a lower median revisit time for the Iridium NEXT constellation.}
\label{fig:iridium_revisit}
\end{figure*}

\subsection{Iridium NEXT Comparison}

To provide a comparison of the results with metrics from actual deployed satellite constellations, this section presents a comparison with the Iridium NEXT constellation. This system consists of a
Walker Star-type constellation with 66 operational satellites deployed in 6 orbital planes, i.e., 11 satellites per plane, with an inclination of 86.4º and an altitude of approximately 780 km. This real constellation has been chosen since it is comparable in terms of the number of satellites to the Walker Delta constellation (64 satellites) presented throughout this paper. To perform the simulations, the same methodology presented at the beginning of this section has been followed, obtaining the orbital parameters of each satellite in TLE format and simulating the positions and dynamics of the satellites using the SGP4 model. Note that up-to-date orbital data for each satellite are available in databases such as CelesTrak.

The first study conducted for Iridium NEXT involves analyzing the probability of establishing visibility with at least one satellite as a function of the minimum elevation angle and the latitude of the user terminal, similar to the analysis shown in Fig.~\ref{fig:sweep_elevacion}(b) for a Walker Delta constellation ($75\degree:64/8/3$) with $h = 1000$~km. Fig.~\ref{fig:iridium_probability}(a) shows the $P(N_{sat} \geq 1)$ in the Iridium NEXT constellation for several latitudes and elevation angles $\epsilon$. According to the probability distribution of visibility, it is practically total for for minimum elevation angles $\epsilon < 15\degree$, for any latitude larger than 55º N. Due to the nearly polar inclination of the orbit, $P(N_{sat} \geq 1)$ increases progressively as polar latitudes are approached. This visibility becomes complete for elevation angles $\epsilon$ below 35º at 90º N. Compared to the previous Walker Delta constellation ($75\degree:64/8/3$), note how Walker Star constellations with polar inclinations tend to provide excellent visibility at Arctic latitudes at the expense of reduced visibility at lower latitudes. To perform a direct comparison between the two constellations, we calculate the difference in terms of visibility probability $\Delta P(N_{sat} \geq 1)$ as $\Delta P(N_{sat} \geq 1) = P(N_{sat_{Walker Delta}} \geq 1) - P(N_{sat_{Iridium NEXT}} \geq 1)$, where $P(N_{sat_{Walker Delta}} \geq 1)$ and $P(N_{sat_{Iridium}} \geq 1)$ correspond to the probability of visibility with at least one satellite from the Walker Delta and Iridium constellations, respectively. According to this definition, positive values of $\Delta P(N_{sat} \geq 1)$ indicate a higher probability of visibility for the Walker Delta constellation compared to Iridium NEXT at a given latitude and minimum elevation angle $\epsilon$, while negative values indicate a higher probability of visibility for the Iridium NEXT constellation compared to Walker Delta. Fig.~\ref{fig:iridium_probability}(b) compares the visibility probability of Iridium NEXT and Walker Delta ($75\degree:64/8/3$) at $h = 1000$~km using $\Delta P(N_{sat} \geq 1)$. Analyzing the comparison, the Walker Delta constellation provides better coverage at latitudes below 75º N, while Iridium NEXT provides better coverage at latitudes above 75º N for intermediate elevation angles in the approximate range $\epsilon \in [20\degree,\; 70\degree]$. Given that both constellations have a very similar number of satellites (64 for Walker Delta vs. 66 for Iridium NEXT), this difference in visibility is primarily explained by the inclination (75º for Walker Delta vs. 86.4º for Iridium NEXT), which causes Iridium to strongly improve the visibility in the Arctic region. Similarly, the Walker Delta design with $i = 75\degree$ distributes this visibility more uniformly across mid-latitudes between 55º N and 75º N. Note that for $\epsilon < 20\degree$, there are hardly any noticeable differences because coverage tends to be complete for both constellations. In contrast, for $\epsilon > 70\degree$, the minimum elevation angle becomes too restrictive, resulting in very limited visibility for both constellations.

Following the analysis of Iridium NEXT, a study and comparison of revisit times is also conducted using the same methodology explained earlier (see Fig.~\ref{fig:revisit_time_elevation}(a)). Fig.~\ref{fig:iridium_revisit}(a) shows the median revisit time $\tau$ for several latitudes and elevation angles $\epsilon$ for the Iridium NEXT constellation. Similar to what was observed in the visibility analysis, the Walker Star-type constellation tends to prioritize coverage at higher latitudes, which is why the median revisit time decreases as latitudes approach 90º N. For instance, for a minimum elevation angle of $\epsilon = 40\degree$, the median revisit time changes from about 6 minutes at a latitude of 55º N to less than 1 minute at a latitude of 90º N. To perform a direct comparison between constellations, $\Delta\,\textrm{Revisit}\,\textrm{Time}$ is defined as $\Delta\,\textrm{Revisit}\,\textrm{Time} = \textrm{Median}\,\textrm{Revisit}\,\textrm{Time}_{IridiumNEXT} - \textrm{Median}\,\textrm{Revisit}\,\textrm{Time}_{WalkerDelta}$, which is shown in Fig.~\ref{fig:iridium_revisit}(b). Thus, positive values, i.e., blue tones in Fig.~\ref{fig:iridium_revisit}(b), indicate a shorter median revisit time for the Walker Delta constellation (75º: 64/8/3), while negative values, i.e., red tones in Fig.~\ref{fig:iridium_revisit}(b), indicate a shorter median revisit time for the Iridium NEXT constellation. Note that this metric has been calculated exclusively in regions with intermittent coverage from both constellations, as this is the domain where the median revisit time can be computed for both constellations. The remaining regions have been marked to indicate the visibility of each constellation. In line with the conclusions previously drawn from the visibility comparison, the average revisit time tends to be shorter for Iridium NEXT at latitudes above 75º N, whereas the average revisit time tends to be shorter in the Walker Delta constellation $(75\degree : 64/8/3)$ when considering latitudes below 75º N. In conclusion, based on the previous analysis Iridium NEXT constellation design is a better choice in terms of visibility and revisit time for user terminals located in purely Arctic regions, such as Svalbard or the northern half of Greenland, due to its near-polar inclination. In contrast, a constellation design like Walker Delta $(75\degree : 64/8/3)$ is better suited for intermediate regions such as the North Atlantic, including land areas such as the Scandinavian countries, Iceland, or the southern half of Greenland.

\subsection{Scope and Limitations}

The methodology presented in this paper focuses on the geometric design of the orbital parameters of a Walker Delta-type constellation in LEO orbits. More specifically, the analysis has focused on how the independent variation of parameters such as orbital inclination, altitude, minimum elevation angle, and satellite footprint affects geometric visibility metrics. Thus, the methodology followed is intended as a geometric design phase, in which the geometric feasibility of a constellation configuration is evaluated before addressing aspects of communication system implementation. Consequently, the recommendations presented should be understood as guidelines for geometric design rather than as a complete methodology for designing a LEO communication network.

Beyond the geometric analysis considered in the work, there are several other factors that influence end-user performance from the perspective of a communications system. For instance, user traffic distribution, resource allocation, scheduling policies, and handover mechanisms determine how satellite visibility translates into the capacity and continuity of service in a satellite communications system. Similarly, beyond free-space path loss, it is necessary to take into account factors such as realistic antenna radiation patterns, beamforming strategies, Doppler effect, adaptive modulation and coding, and link-layer protocols, all of which depend on the selected communication architecture and hardware implementation.

From a network perspective, the study assumes that a visible satellite can potentially provide service to a user terminal on the ground. However, for communications requiring Internet connectivity, factors such as gateway availability, which must also have visibility towards the satellite, inter-satellite links (ISLs), or routing strategies must be taken into account, as these determine how traffic is transported from the user terminal towards the Internet via the constellation. Again, these aspects depend on the network architecture adopted by operators and have a significant influence on end-to-end latency, throughput, and service availability. Finally, from an economic and satellite deployment perspective, there are considerations such as launch strategy and cost, as well as satellite manufacturing costs. These factors must also be taken into account during the engineering design of operational constellations.

In summary, the resulting geometric recommendations should subsequently be combined with communication-system simulations and optimization frameworks that account for the additional implementation constraints described above.

\section{Conclusions}

This study has provided an overview of the current deployment of constellations in LEO orbits and their characteristics, and it has analyzed the main geometric and propagation-related factors that guide coverage performance in LEO satellite constellations, with particular emphasis on regional coverage over the North Atlantic. Through analytical modeling and orbital simulations, the study has conducted a parametric analysis to evaluate the individual influence of key constellation parameters such as minimum elevation angle, orbital inclination, altitude, and satellite footprint, and what their impact is on visibility probability, revisit time, and propagation losses.

The results have demonstrated that elevation angles closer to the horizon and higher altitudes maximize the coverage area at the expense of greater signal loss due to link range and increasing atmospheric attenuation, which is particularly noticeable in the millimeter-wave range. In contrast, higher elevation angles substantially reduce coverage capability due to the shrinking footprint, which can potentially limit satellite visibility at high latitudes. Therefore, this analysis shows the strong trade-off between coverage extension and link quality when selecting the minimum elevation angle. Similarly, the analysis further reveals the effect of elevation angle on the probability of visibility and revisit time between ground stations and satellites for multiple latitudes, thereby providing insights into the coverage offered by LEO constellations as a function of elevation angle across the North Atlantic. 

Finally, the effect of the inclination angle of a LEO constellation on coverage across the North Atlantic has been analyzed, demonstrating that the constellation inclination is one of the dominant design parameters for regional high-latitude deployments. Medium-size Walker Delta \mbox{$(i:64/8/3)$} constellations with inclinations below approximately 70$\degree$ cannot provide robust visibility across the North Atlantic, while larger inclinations are required to ensure visibility continuity in North Atlantic and Arctic regions for $40\degree$ minimum elevation angles and 1000 km altitude. Moreover, the simulations show that visibility probability is maximized at latitudes close to the orbital inclination, highlighting the importance of designing constellation geometries customized to the target service area instead of relying exclusively on either low-inclination or fully polar deployments. In addition, for comparison purposes, several of the simulated parameters were compared against those of the real and operational Iridium NEXT constellation to provide a reference for the obtained results.

Overall, the presented numerical results, along with their qualitative analysis, provide practical design guidelines for regional LEO constellation engineering, supporting future NTN deployments targeting maritime, aviation, Arctic, and remote high-latitude connectivity scenarios across the North Atlantic.

\bibliographystyle{IEEEtran} 
\bibliography{refs}

\end{document}